\documentclass[12pt]{article}
\usepackage{epsfig,amsfonts,amssymb,amsmath}
\usepackage{hyperref}
\usepackage{cite}
\input epsf.sty
\newcommand{\m}{m}

\def\ZZZ{{\hbox{ Z\kern-1.6mm Z}}}
\def\RRR{{\hbox{ R\kern-2.4mm R}}}
\def\CCC{{\hbox{ C\kern-2.0mm C}}}
\def\zzz{{\hbox{z\kern-1mm z}}}

\newcommand{\f}{\frac}

\newcommand{\qeq}{{\hbox{=\kern-2.3mm ? \kern.5mm }}}
\renewcommand{\qeq}{=}

\newcommand{\non}{\nonumber}
\newcommand{\be}{\begin{eqnarray}}
\newcommand{\ee}{\end{eqnarray}}
\newcommand{\ben}{\begin{eqnarray}\displaystyle}
\newcommand{\een}{\end{eqnarray}}

\newcommand{\p}{\partial}
\newcommand{\sectiono}[1]{\section{#1}\setcounter{equation}{0}}

\def\one{{\hbox{ 1\kern-.8mm l}}}
\def\zero{{\hbox{ 0\kern-1.5mm 0}}}

\newcommand{\bea}[1]{\begin{eqnarray}\label{#1} }
\newcommand{\eea}{\end{eqnarray}}

\usepackage{bm}
\usepackage[table]{xcolor}

\begin{document}

\begin{flushright}
HRI/ST/1704 
\end{flushright}

\baselineskip 24pt

\begin{center}
{\Large \bf  Theta Expansion of First Massive Vertex Operator in Pure Spinor}

\end{center}

\vskip .6cm
\medskip

\vspace*{4.0ex}

\baselineskip=18pt

\begin{center}

{\large 
\rm Subhroneel Chakrabarti$^{a,b}$, Sitender Pratap Kashyap$^{a,b}$, Mritunjay Verma$^{a,b,c}$ }

\end{center}

\vspace*{4.0ex}

\centerline{\it \small $^a$Harish-Chandra Research Institute,
Chhatnag Road, Jhunsi,
Allahabad 211019, India}
\centerline{\it \small $^b$Homi Bhabha National Institute, 
Anushakti Nagar,
    Mumbai 400085, India}
\centerline{ \it \small $^c$International Centre for Theoretical Sciences,
 Hesarghatta,
Bengaluru 560089, India.}

\vspace*{1.0ex}
\centerline{\small E-mail: subhroneelchak,
sitenderpratap, mritunjayverma@hri.res.in}

\vspace*{5.0ex}

\centerline{\bf Abstract} \bigskip
We provide the covariant superspace equations that are sufficient to determine the complete $\theta$ expansion of the vertex operator of the open string massive states with $(mass)^2=1/\alpha'$ in pure spinor formalism of superstring theory. These equations get rid of the redundant degrees of freedom in superfields and are consistent with the BRST conditions derived in \cite{Berkovits1}. Further, we give the explicit $\theta$ expansion of the superfields appearing in the unintegrated vertex to 
$O(\theta^3)$. Finally, we compute the contribution  to a 3-point tree amplitude with the resulting vertex operator upto $O(\theta^3)$ and find its
 kinematic structure to be identical to the corresponding RNS computation.

\vfill


\vfill \eject

\baselineskip 18pt

\tableofcontents

\sectiono{Introduction } 
In string theory, some of the open string excitations can be interpreted as gauge particles. After some suitable compactification
of the ten dimensional theory involving the open strings ending on D-brane, one expects to get the elementary gauge particles we see in the four dimensional world
around us. The effect of massive stringy states is expected to be relevant as correction to standard model at energy scales higher than that of the scale at which 
standard model is known to be accurate.

\vspace*{.06in}In this paper, we focus on the first massive open string states in 10 dimensional theory. They consist of 128 bosonic and 128 fermionic degrees of 
freedom forming a spin two massive supermultiplet of 10 dimensional $\mathcal{N}=1$ supersymmetry. The bosonic degrees of freedom consist of a symmetric-traceless
spin 2 field $g_{mn}$ having 44 independent components and a 3-form field $b_{mnp}$ having 84 independent components. The fermionic degrees of freedom consist of a 
spin 3/2 field $\psi_{m\alpha}$ having 128 degrees of freedom (see, e.g. \cite{witten}).  

\vspace*{.06in}To describe the above supermultiplet we shall be using the pure spinor formalism.  As is well known, the pure spinor formalism is a manifestly 
super-poincare covariant formalism \cite{Berkovits, Berkovits7} (for review, see \cite{Berkovits2, Berkovits12, Mafra1, Oliver1, Joost1}). Due to the feature that it keeps all the spacetime symmetries manifest, the pure spinor formalism 
provides a very efficient method for computing the scattering amplitudes \cite{BerkovitsNekrasov, Berkovits6, Berkovits7, Berkovits5, Berkovits9, Berkovits_Mafra}. The equivalence of the pure spinor formalism with RNS and Green-Schwarz formalism has been shown at the level of the BRST cohomology and also
through quite a few explicit amplitude computations \cite{ Berkovits3, Berkovits4, Berkovits8, Berkovits10}. 

\vspace*{.06in} The physical states in the pure spinor formalism are described as the states in the cohomology of a BRST operator constructed with the help of a 
pure spinor field. The vertex operators describing these physical string states are defined in a manifestly super-poincare covariant manner in terms of appropriate 
superfields in $d=10$ superspace. The vertex operator for the massless states was given in \cite{Berkovits}. In \cite{Berkovits1}, it was shown explicitly 
that the pure spinor BRST cohomology includes the states which appear at first massive level of open string as described above. The unintegrated form of the vertex 
operator for these states was also constructed in \cite{Berkovits1}. 

\vspace*{.06in}One drawback of working in the superspace is that the superfields contain much more degrees of freedom than the actual physical degrees of freedom.
Hence, before doing any calculation using a superfield, one needs to express all the coefficients of the superfield in its $\theta$ expansion only in terms of the physical
fields. This typically requires some gauge fixing and the knowledge of  \textit{algebraic} and \textit{differential} relations\footnote{By algebraic relations we mean relations that do not involve supercovariant derivative and by differential relations we mean those relations which involve one supercovariant derivative.} between the superfields. The $\theta$ expansion for the massless vertex
operator was done in \cite{Harnad, Ooguri, Policastro}. 

\vspace*{.06in}As mentioned above, the $\theta$ expansion typically requires some relations between the superfields. In the case of first massive open string states, 
3 basic superfields are $B_{mnp}, G_{mn}$ and $\Psi_{m\alpha}$, whose lowest components are $b_{mnp}, g_{mn}$ and $\psi_{m\alpha}$ respectively. The vertex operator
can be expressed fully in terms of any one of these superfields. For example, the vertex operator constructed in \cite{Berkovits1} was expressed solely in terms of the 
single superfield $B_{mnp}$. It turns out that complete $\theta$ expansion requires differential relations among these 3 superfields in addition to the algebraic relations already given in \cite{Berkovits1} in the rest frame.  

\vspace*{.06in}One of these differential relationships between the superfields $B_{mnp}$ and $\Psi_{m\alpha}$ was given in \cite{Berkovits1}. But, this was given in the rest frame 
(equation \eqref{D_Bmnp_rest} in this paper). However, the covariant result for the differential relations is necessary when there are more than one massive states 
present in an amplitude\footnote{If there is only one massive state and 
the rest are massless states, one can go to the rest frame of the massive state and use the expression of massive vertex operator in the rest frame.}. We provide the 
covariant version of this relation, along with the other differential relations between the three superfields mentioned above. These are very crucial and 
sufficient to determine the $\theta$ expansion to all orders in a covariant manner. 

\vspace*{.06in}Rest of the paper is organized as follows. In section \ref{sec:review}, we briefly review some elements of the pure spinor formalism as needed in our
analysis. After giving some general results in subsection \ref{subsec:general_result}, we review the construction of unintegrated vertex operator for the first
massive states in subsection \ref{consverop}. In section \ref{ingredients}, we give the differential relations between the superfields $B_{mnp}, G_{mn}$ and
$\Psi_{m\alpha}$. The equations \eqref{D_Gmn1}-\eqref{cons_theta=0} are our main results. These give us the BRST invariant vertex operator and allow us to systematically perform the theta expansion. 
In section \ref{thetaex}, we give the procedure to determine the higher $\theta$ components using the results of section \ref{ingredients}. We also give the $\theta$ expansion for the superfields $\Psi_{m\alpha}, G_{mn}$ and $B_{mnp}$ upto order $\theta^3$. The expression for order more than 3 can be worked out straightforwardly whenever needed. In section \ref{amplical}, we consider an illustrative calculation of 3-point tree amplitude involving one massive state, compute the contribution to the amplitude resulting from the massive vertex operator upto $O(\theta^3)$  and find the kinematic structure to be same as that computed using RNS formalism. Finally, we conclude with discussion in section \ref{discus}. The appendices elaborate on our notations and conventions and contain some useful identities used in the paper. 

 \vspace*{.06in}Before moving ahead, we should mention that the $\theta$ expansion and the scattering of massive states in pure spinor formalism were also considered
 in \cite{Park}.  The $\theta$ expansion was done with the help of a covariant version of the relation between $B_{mnp}$ and $\Psi_{m\alpha}$ superfields  in the rest
 frame as given in \cite{Berkovits1}. However, the covariant expression used by \cite{Park} is in conflict with another equation of \cite{Berkovits1} as we mention in 
 footnotes \ref{f1}, \ref{f2}, \ref{f3} and appendix \ref{sec:check} of this paper. Moreover, the analysis of \cite{Park} could not have been complete, even with the correct covariant generalization of the 
 relation mentioned above. This is because the additional differential relations \eqref{D_Gmn1} and \eqref{D_Psi1} given in section \ref{ingredients} are indispensable for $\theta$ expansion.

\section{Brief review of pure spinor formalism }
\label{sec:review}
In this section, we briefly recall some of the results of the minimal pure spinor formalism. For details, we refer the reader to original papers. The purpose of this section is to mainly introduce the fields which are used in the formalism. After this, we shall recall some of the details of the first massive vertex operator \cite{Berkovits1} which are
relevant for our purpose.  
\subsection{Some general results }
\label{subsec:general_result}

\vspace*{.06in}As mentioned earlier, the pure spinor formalism is a formalism to quantize superstrings covariantly. Unlike RNS and Green-Schwarz formalisms, all the underlying
symmetries, namely Poincare and supersymmetry remain manifest in this formalism. Restricting to open strings, the world-sheet CFT in the flat spacetime is described by the action
\be
S=\f{2}{\alpha'}\int d^2z \left(\f{1}{2}\p X^m\bar\p X_m+p_\alpha\bar\p\theta^\alpha-w_\alpha\bar\p\lambda^\alpha\right)
\ee
where, $m=0,1,,\cdots,9$ and $\alpha=1,\cdots,16$. 

\vspace*{.06in}The conformal weights of the fields $p_\alpha, w_\alpha,\theta^\alpha$ and $\lambda^\alpha$ are $1,1,0,0$ respectively. Moreover, the field
$p_\alpha$ is a left handed Majorana-Weyl spinor whereas $\theta^\alpha$ is right handed Majorana Weyl spinor\footnote{All the left-handed spinors carry 
lower spinor indices whereas all the right-handed spinors carry upper spinor indices.}. The fields $w_\alpha$ and $\lambda^\alpha$ are bosonic objects which transform as left and right handed
Majorana Weyl spinor respectively under the Lorentz transformation (hence, violating the spin-statistics theorem). The field $\lambda^\alpha$ satisfies an important constraint, the so called
pure spinor constraint
\be
\lambda^\alpha\gamma^{m}_{\alpha\beta}\lambda^\beta=0 
\ee
where, $\gamma^{m}$ are the $16\times 16$ gamma matrices, described in detail in the appendix \ref{identi}.

\vspace*{.06in}The ghost Lorentz and ghost number currents $N^{mn}$ and $J$ are given by
\be
N^{mn}=\f{1}{2}w_\alpha (\gamma^{mn})^\alpha_{\;\;\beta}\lambda^\beta\quad,\qquad
J=w_\alpha\lambda^\alpha
\ee
The physical spectrum of the theory corresponds to the cohomology of the following BRST operator
\be
Q=\oint dz\ \lambda^\alpha(z) d_\alpha(z)
\ee
which generates the following transformations
\be
\delta X^m=\lambda\gamma^m\theta\quad,\quad \delta\theta^\alpha=\lambda^\alpha\quad,\quad \delta \lambda^\alpha=0\quad,\quad \delta d_\alpha=-\Pi^m(\gamma_m\lambda)_\alpha\quad,\quad \delta w_\alpha=d_\alpha
\ee
where, $d_\alpha$ and $\Pi^m$ are supersymmetric invariant combinations 
\be
d_\alpha=p_\alpha-\f{1}{2}\gamma^m_{\;\;\alpha\beta}\theta^\beta\partial X_m-\f{1}{8}\gamma^m_{\alpha\beta}\gamma_{m\sigma\delta}\theta^\beta\theta^{\sigma}\partial\theta^\delta
\ee
\be
\Pi^m=\partial X^m+\f{1}{2}\gamma^m_{\alpha\beta}\theta^\alpha\partial\theta^\beta
\ee
The OPE between various objects is given by
\be
d_\alpha(z)d_\beta(w)=-\f{\alpha'\gamma^m_{\alpha\beta}}{2(z-w)}\Pi_m(w)+\cdots
\quad,\qquad
d_\alpha(z)\Pi^m(w)=\f{\alpha'\gamma^m_{\alpha\beta}}{2(z-w)}\partial\theta^\beta(w)+\cdots\non
\ee
\be
d_\alpha(z)V(w)=\f{\alpha'}{2(z-w)}D_\alpha V(w)+\cdots \quad,\qquad
\Pi^m(z)V(w)=-\f{\alpha'}{(z-w)}\partial^mV(w)+\cdots\non
\ee
\be
 \Pi^m(z)\Pi^n(w)=-\f{\alpha'\eta^{mn}}{(z-w)^2}+\cdots\quad,\qquad N^{mn}(z)\lambda^\alpha(w)=\f{\alpha'(\gamma^{mn})^\alpha_{\;\;\beta}}{4(z-w)}\ \lambda^\beta(w)+\cdots
\non
\ee
\be
N^{mn}(z)N^{pq}(w)=-\f{3(\alpha')^2}{4(z-w)^2}\eta^{m[q}\eta^{p]n}+\f{\alpha'}{2(z-w)}\Bigl(\eta^{p[n}N^{m]q}-\eta^{q[n}N^{m]p}\Bigl)+\cdots\non
\ee
\be
J(z)J(w)=-\f{(\alpha')^2}{(z-w)^2}+\cdots\quad,\qquad  J(z)\lambda^\alpha(w)=\f{\alpha'}{2(z-w)}\lambda^\alpha(w)+\cdots\label{OPE_eq}
\ee
where, $V$ is an arbitrary superfield, $\p$ denotes the derivative with respect to world-sheet coordinate, $\p_m$ denotes the derivative with respect to the spacetime coordinate $X^m$ and
 \be D_\alpha\equiv\p_\alpha+\gamma^m_{\alpha\beta}\theta^\beta\p_m
\ee
 denotes the supercovariant derivative. The
$\cdots$ terms denote the non-singular terms.

\vspace*{.06in}The scattering of N external string states at tree level is described by the amplitude
\be
\mathcal{A}_N=\langle V^1V^2V^3\int U^4\cdots\int U^N \rangle	 \label{treeamp}
\ee                       
 $V$ and $U$ in the above expression denote the unintegrated and integrated vertex operators respectively. The correlation functions of pure spinor fields are computed using the OPEs given in equation \eqref{OPE_eq} and the identities of appendix \ref{PureSuperspace}\footnote{In appendix \ref{PureSuperspace}, we only give the identities which are used in this paper. See \cite{Mafra1} for a complete list.}. 

\vspace*{.06in}After this brief recollection of general pure spinor results, we now turn to unintegrated massive vertex operator.

\subsection{Unintegrated massive vertex operator at $(mass)^2=\f{1}{\alpha'}$  }
 \label{consverop}
In this subsection, we focus on the open string states at first mass level, i.e. $m^2=\f{1}{\alpha'}$. The unintegrated vertex operator for these states was constructed in \cite{Berkovits1}. We review this construction below. As mentioned in the introduction, at the first mass level, the open string spectrum comprises of 128 bosonic and 128 fermionic degrees of freedom contained in a traceless symmetric tensor $g_{mn}$, a three-form field $b_{mnp}$ and a spin-$3/2$ field $\psi_{m\alpha}$. These fields satisfy the following constraints
\be 
\eta^{mn}g_{mn}=0\quad ;\quad  \p^m g_{mn}=0 \quad ;\quad \p^m b_{mnp}=0 \quad ;\quad \p^m\psi_{m\alpha}=0 \quad ; \quad \gamma^{m\alpha\beta} \psi_{m\beta}=0 
\label{constraints_theta_ind_comp}
\ee
Due to these constraints, the number of independent components in $g_{mn}, b_{mnp}$ and $\psi_{m\alpha}$ is $44, 84$ and $128$ respectively. Further, these form
a massive spin-2 supermultiplet in 10 dimensions.

\vspace*{.06in}The unintegrated vertex operator describing the physical states at mass level $n$, i.e., $m^{2}=\f{n}{\alpha'}$ is constructed out of objects\footnote{These objects are constructed using $\Pi^m, \p\theta^\alpha,d_\alpha, \lambda^\alpha, J \textup{ and } N^{mn}$.} with ghost number $1$ and conformal dimension $n$. Consequently, the most general 
unintegrated vertex operator at first massive level ($n=1$) of the open string can be written as 
\ben
V&=&\partial \lambda^\alpha A_\alpha(X,\theta)+:\partial \theta^\beta \lambda^\alpha B_{\alpha\beta}(X,\theta): 
+:d_\beta\lambda^\alpha C^\beta_{\;\alpha}(X,\theta):+:\Pi^m\lambda^\alpha H_{m\alpha}(X,\theta):
\nonumber\\
&&\;\;\;+:J\lambda^\alpha E_\alpha(X,\theta):+:N^{mn}\lambda^\alpha F_{\alpha mn}(X,\theta):                                                    \label{vertex_ansatz}
\een
where $A_\alpha, B_{\alpha\beta}, C^{\beta}_{\;\;\alpha}, H_{m\alpha},E_{\alpha}$ and $F_{\alpha mn}$ are general superfields, unconstrained as of now. In accordance with \cite{Berkovits1}, the normal ordering $:\;:$ is defined as follows
\be 
:AB:(z) \equiv \frac{1}{2 \pi i} \oint_z  \frac{dw}{w-z} A(w)B(z)  
\ee
 where, $A$ and $B$ are any two operators.

\vspace*{.06in}The equation
of motion for the superfields in \eqref{vertex_ansatz} is determined by the on-shell condition $QV=0$ which yields the following set of equations\footnote{Note that some of the numerical
factors in various expressions in this paper will not agree with that in \cite{Berkovits1} since we are using different convention for (anti)symmetrization. See appendix \ref{saconv} for details.}
\ben
(\gamma_{mnpqr})^{\alpha\beta} \left[D_\alpha B_{\beta\sigma}-\gamma^s_{\alpha\sigma}H_{s\beta}\right]=0\label{1st}
\ee
\be
(\gamma_{mnpqr})^{\alpha\beta} \left[D_\alpha H_{s\beta}-\gamma_{s\alpha\sigma}C^\sigma_{\;\;\beta}\right]=0\label{2nd}
\ee
\be
(\gamma_{mnpqr})^{\alpha\beta} \left[D_\alpha C^\sigma_{\;\;\beta}+\delta^\sigma_{\;\;\alpha}E_\beta+\f{1}{2}(\gamma^{st})^\sigma_{\;\;\alpha}F_{\beta st}\right]=0
\label{Berkovits_eqns}
\ee
\begin{eqnarray}
&&(\gamma_{mnpqr})^{\alpha\beta} \Bigl[D_\alpha A_{\beta}+B_{\alpha\beta}+\alpha'\gamma^s_{\beta\sigma}\p_s C^\sigma_{\;\;\alpha}-\f{\alpha'}{2}D_\beta E_\alpha+\f{\alpha'}{4}(\gamma^{st}D)_{\beta}F_{\alpha st}\Bigl]\nonumber\\
&&\hspace*{1in}=2\alpha'\gamma_{mnpqr}^{\alpha\beta}\gamma_{\alpha\beta}^{vwxys}\eta_{st}K^t_{\;\;vwxy} 
\end{eqnarray}
\begin{eqnarray}
&&(\gamma_{mnp})^{\alpha\beta} \Bigl[D_\alpha A_{\beta}+B_{\alpha\beta}+\alpha'\gamma^s_{\beta\sigma}\p_s C^\sigma_{\;\;\alpha}-\f{\alpha'}{2}D_\beta E_\alpha+\f{\alpha'}{4}(\gamma^{st}D)_{\beta}F_{\alpha st}\Bigl]\nonumber\\
&&\hspace*{1in}=16\alpha'\gamma_{mnp}^{\alpha\beta}\gamma_{\alpha\beta}^{wxy}\eta_{st}K^s_{\;\;wxys} \label{massiveKG}
\end{eqnarray}
\begin{eqnarray}
(\gamma_{mnpqr})^{\alpha\beta} D_\alpha E_{\beta}=(\gamma_{mnpqr}\gamma^{vwxy}\gamma_s)^{\alpha}_{\;\;\alpha}K^s_{\;\;vwxy}
\end{eqnarray}
\begin{eqnarray}
(\gamma_{mnpqr})^{\alpha\beta} D_\alpha F^{st}_{\beta}=-2(\gamma_{mnpqr}\gamma^{vwxy}\gamma^{[s})^{\alpha}_{\;\;\alpha}K^{t]}_{\;\;vwxy}\label{7th}
\end{eqnarray}
where, $K^{s}_{\;\;vwxy}$ are arbitrary superfields satisfying
\ben 
&&-2:N_{st}\lambda^\alpha\lambda^\beta:(\gamma^{vwxy}\gamma^{[s})_{\alpha\beta}K^{t]}_{\;\;vwxy} 
+ :J\lambda^\alpha\lambda^\beta:(\gamma^{vwxy}\gamma_{s})_{\alpha\beta}K^{s}_{\;\;vwxy}\non\\
&&\hspace{.4in}+\alpha'\lambda^\alpha\p\lambda^\beta\Big[2(\gamma^{vwxys})_{\alpha\beta}\eta_{st}K^t_{\;\;vwxy}+16(\gamma^{wxy})_{\alpha\beta}K^s_{\;\;wxys}\Big]=0
\een
which follows from the identity 
\be
:N_{st}\lambda^\alpha\lambda^\beta:\gamma^s_{\beta\gamma}-\f{1}{2}:J\lambda^\alpha\lambda^\beta:(\gamma_{t})_{\beta\gamma}
=\f{5\alpha'}{4}\lambda^\alpha\p\lambda^\beta(\gamma_t)_{\beta\gamma}-\f{\alpha'}{4}\lambda^\delta\p\lambda^\beta(\gamma_{st})^\alpha_{\;\;\delta}(\gamma^s)_{\beta\gamma}
\ee
Due to the nilpotency of the BRST operator $Q$, any vertex operator $V$ enjoys a gauge freedom given by the transformation
\be V(z) \rightarrow V(z) +Q\Omega(z)														\label{gauge_trans}
\ee
This gauge freedom can be used to impose the following algebraic conditions on the superfields
\be 
B_{\alpha\beta}&=&(\gamma^{mnp})_{\alpha\beta}B_{mnp}\quad\;, \qquad C^{\alpha}_{\;\;\beta}=(\gamma^{mnpq})^\alpha_{\;\;\beta}C_{mnpq}\non
\ee
\be
\gamma^{m\alpha\beta}F_{\beta mn}&=&0\quad\;, \qquad\qquad (\gamma^m H_m)_{\alpha}=0\label{solution_berkovits}
\ee

Using these and the equations of motion \eqref{1st} - \eqref{7th}, all the superfields can be solely expressed in terms of the single superfield $B_{mnp}$ as  
\be 
&&H_{s\alpha}=\f{3}{7}(\gamma^{mn})_{\alpha}^{\;\;\beta}D_{\beta}B_{mns}\;, \quad C_{mnpq}=\f{1}{2}\p_{[m}B_{npq]} \;, \quad E_{\alpha}=0=A_\alpha \non\\[.2in]
&&\hspace{.8 in}F_{\alpha mn} =\f{1}{8}\biggl(7\p_{[m}H_{n]\alpha} + \p^q(\gamma_{q[m})_\alpha^{\;\;\beta}H_{n]\beta}\biggl)
\label{2.14}
\ee
Further, one also finds 
\be
K^m_{\;\;\;npqr}&=& \frac{1}{1920}\left[(\gamma_{npqrs})^{\alpha \beta}D_\alpha F^{ms}_\beta-\frac{1}{3}(\gamma_{su[npq})^{\alpha \beta} \delta^m_{r]} D_\alpha F^{su}_\beta\right] 
\ee
The above solution, when substituted in \eqref{massiveKG}, simplifies to
\be
\big( \partial^2 -\frac{1}{\alpha '} \big)B_{mnp} =0\label{KG11}
\ee
which demonstrates that the states described by $B_{mnp}$ are indeed massive with $(mass)^2=\frac{1}{\alpha '}$. 

\vspace*{0.06in}In order to show that the superfields contain the physical fields $g_{mn}, b_{mnp}$ and $\psi_{m\alpha}$, it is convenient to go to the rest frame $\vec{k}=0$, where $\vec{k}$ denotes
the spatial momenta. In the following, we shall label the spatial indices using the beginning roman alphabets, namely, $a,b,c$ etc.

\vspace*{.06in}As argued in \cite{Berkovits1} using the supersymmetry transformation properties, $B_{mnp}$ and $H_{m\alpha}$ satisfy in the rest frame,
\be 
D_\alpha B^{abc}=12(\gamma^{[ab}\Psi^{c]})_\alpha \quad   ; \qquad B^{0ab}=0							 \label{D_Bmnp_rest}
\ee
and\footnote{In fact the covariant expression of \eqref{D_Bmnp_rest}
as proposed in \cite{Park}, after substituting in $H_{s\beta}$ as given in \eqref{2.14} reduces in the rest frame to $H^b_{\beta}=-96 \Psi^b_{\beta}$. This is in clear contradiction with \eqref{2.20}. \label{f1}}  
\be 
H^c_{\alpha}=-72 \,\Psi^c_\alpha
\label{2.20}
\ee
where, $\Psi^c_\alpha$ is an arbitrary tensor-spinor superfield satisfying 
\be
(\gamma_{a})^{\beta\alpha}\Psi^a_\alpha=0														
\label{psi_constraint_rest}
\ee
The spin-3/2 field $\psi^a_{\alpha}$ is defined to be the $\theta$ independent component of $\Psi^a_{\alpha}$ i.e. 
\be
\psi^a_{\alpha}=\Psi^a_{\alpha}\Big|_{\theta=0}
\ee
Furthermore, the physical fields $g_{ab}$ and $b_{abc}$ are defined to be the $\theta$ independent components of $G_{ab}$ and $B_{a b c}$ respectively i.e. 
\be 
g^{ab}\equiv G^{ab}\Big|_{\theta=0} \quad ; \quad\quad b_{abc}\equiv B_{abc}\Big|_{\theta=0}								
\ee
where, the superfield $G^{ab}$ is defined to be 
\be
G_{ab}\equiv 2\; D_\alpha\gamma_{(a}^{\alpha\beta}\Psi_{b)\beta}	\qquad,\qquad\quad \eta_{ab}G^{ab}=0													
\label{graviton_rest}
\ee
Due to the fact that $b_{abc}$ is anti-symmetric in all the indices and equations \eqref{psi_constraint_rest} and \eqref{graviton_rest},
the fields $b_{abc}, g_{ab}  $ and $\psi_{\alpha}^a$ contain precisely the desired number of degrees of freedoms, namely, $84, 44$ and $128$ respectively. This shows
that in the rest frame we have the correct counting for the degrees of freedom in the massive superfields. For these superfields to represent the massive spin-2 supermultiplet, the higher $\theta$ components of the superfields $B_{mnp}, \Psi_{m\alpha}$ and $G_{mn}$ must be determined completely in terms of $b_{mnp}, g_{mn}$ and $\psi_{m\alpha}$. As we shall see in the next two sections, this indeed turns out to be the case.

\vspace*{.06in}In the next section, we shall write down the covariant generalizations of the rest frame results given above and the differential relations between the superfields $B_{mnp}, G_{mn}$ and $\Psi_{m\alpha}$ as mentioned earlier. These will enable us to do the $\theta$ expansion completely. 
 
\sectiono{Ingredients for $\theta$ expansion}
\label{ingredients}
As mentioned in the introduction, one drawback of working in the superspace formalism is that a given superfield contains much more degrees of freedom than the actual physical degrees of freedom of the theory. In our case also, by looking at the coefficients in the $\theta$ expansion of the superfields $B_{mnp}, G_{mn}$ and $\Psi_{m\alpha}$, we can easily convince ourselves that they contain much more degrees of freedom than 128+128 provided by the physical fields $g_{mn}, b_{mnp}$ and $\psi_{m\alpha}$. Thus, it is imperative that we express the higher $\theta$ components of these superfields in terms of the physical fields thereby removing the redundant degrees of freedom. 

\vspace*{.06in}To ensure that $g_{mn}, b_{mnp}$ and $\psi_{m\alpha}$ are the only physical degrees of freedom, there must be relations expressing $D_\alpha\Psi_{m\beta}$ in terms of $G_{mn}, B_{mnp}$ and $D_\alpha G_{mn}$, $D_\alpha B_{mnp}$ in terms of $\Psi_{m\alpha}$. These will provide the recursive relations\footnote{Note that the gauge invariance \eqref{gauge_trans} has already been completely exploited in writing down the solution \eqref{2.14}.} relating the higher $\theta$ components of the superfields to the lowest components $g_{mn}, b_{mnp}$ and $\psi_{m\alpha}$.  Along with these, the algebraic constraints such as $k^mB_{mnp}=0$ are also needed to remove the extra degrees of freedom at the zeroth order in $\theta$ expansion. We need to ensure that all these relations are consistent with the on-shell condition $QV=0$ (or equivalently equations\eqref{1st} to \eqref{7th}). 

\vspace*{.06in}In this section, we give the above mentioned relationships among the superfields. In the process, we also give the covariant generalizations of the rest frame results \eqref{D_Bmnp_rest} - \eqref{graviton_rest} given in section \ref{consverop}. 
We shall be very brief and just state the result. One can check the validity of these by writing them in the rest frame and verifying that they agree with those in
the subsection \ref{consverop} and satisfy all the equations. In appendix \ref{sec:check}, we indicate how to check this systematically. For simplicity, we work in the momentum space in what follows\footnote{i.e. we replace all the $\p_m $ by $ i \,k_m .$}.

\vspace*{.06in}We start by recalling the rest frame result \eqref{2.20} which relates $\Psi^b_\beta$ and $H^b_\beta$. We also note that $\Psi^0_\beta$ does not
appear anywhere in the rest frame analysis. Hence, we can set it to zero in the rest frame. In fact, a non zero $\psi^0_\beta$ will not be consistent
with the fact that $\psi_{s\beta}$ contains $128$ degrees of freedom in the rest frame unless further constraints are imposed on $\Psi^a_{\beta}$. This means that the covariant generalizations of \eqref{2.20} and \eqref{psi_constraint_rest} can be taken to be
\be
H_{m\beta}=-72\Psi_{m\beta}\quad,\qquad\qquad (\gamma_{m})^{\alpha\beta}\Psi^m_{\beta}\;\;=\;\;0
\label{3.21}
\ee
We also need the covariant generalization of \eqref{graviton_rest} which is given by
\be
G_{mn}=2D_\alpha\gamma^{\alpha\beta}_{(m}\Psi_{n)\beta}\quad,\qquad \quad \eta^{mn}G_{mn}=0
\label{3.25}
\ee
These results have the correct limit in the rest frame. 

\vspace*{.06in}We now write down the relations between the various superfields and all the necessary constraints which are needed to ensure that superfields contain only the physical degrees of freedom. 
\be
D_{\alpha}G_{sm} = 16i k^{p}(\gamma_{p(s}\Psi_{m)})_\alpha \label{D_Gmn1}
\ee
\be
D_\alpha B_{mnp}=12 (\gamma_{[mn}\Psi_{p]})_\alpha + 24\alpha' k^t k_{[m}(\gamma_{|t|n}\Psi_{p]})_\alpha                             \label{D_Bmnp_general}
\ee
\be
D_{\alpha}\Psi_{s\beta}= \f{1}{16} G_{sm}\gamma^{m}_{\alpha\beta}+\f{i}{24}k_mB_{nps}(\gamma^{mnp})_{\alpha\beta}
-\f{i}{144}k^mB^{npq}(\gamma_{smnpq})_{\alpha\beta} \label{D_Psi1}
\ee
\be
 (\gamma^{m})^{\alpha \beta}\Psi_{s\beta}=0\quad ; \quad k^m \Psi_{m\beta}=0\quad ; \quad  k^m B_{mnp}=0 \quad ; \quad  k^m G_{m n}=0 \;\; \& \;\; \eta^{mn}G_{mn}=0\label{cons_theta=0}
\ee
In order to arrive at the above results\footnote{The equation 67 of \cite{Park} is to be contrasted with the equation \eqref{D_Bmnp_general} given above. Note that the term proportional to $\alpha'$ is not present in \cite{Park}. \label{f2}}, we can write an ansatz with arbitrary coefficients using the superfields and momentum vector 
$k^m$. The coefficients can be fixed by demanding that the ansatz reduces to the rest frame result and it is consistent with other equations in which it appears. For instance, to obtain the relation \eqref{D_Psi1}, we expand $D_\alpha\Psi_{s\beta}$ in the basis of linearly independent gamma matrices using the identity \eqref{bispinor_expansion}. The superfields appearing in the expansion can be fixed by using the equations of motion $QV=0$ and demanding its consistency with the solution \eqref{2.14}. Similarly, the coefficients appearing in the ansatz for \eqref{D_Gmn1} and \eqref{D_Bmnp_general} can be fixed by using equations \eqref{2.14}, \eqref{3.21}, \eqref{3.25} and \eqref{D_Psi1}. 

\vspace*{.06in}The constraints given in \eqref{cons_theta=0} are necessary to ensure that the lowest components of the superfields contain only the physical degrees of freedom. To see this, we note that these conditions, on the $\theta$ independent components of superfields, imply
\be
 (\gamma^{m})^{\alpha \beta}\psi_{s\beta}=0\quad ; \quad k^m \psi_{m\beta}=0\quad ; \quad  k^m b_{mnp}=0 \quad ; \quad  k^m g_{m n}=0 \;\; \& \;\; \eta^{mn}g_{mn}=0
\ee
These conditions are the momentum space version of the constraints given in \eqref{constraints_theta_ind_comp} and guarantee that $\psi_{s\beta}, g_{mn}$ and $b_{mnp}$ have the correct number of degrees of freedom, namely 128, 44 and 84 respectively. 

\vspace*{.06in}Once the equations \eqref{D_Gmn1} - \eqref{cons_theta=0} and \eqref{solution_berkovits} - \eqref{KG11} hold, all the equations resulting from $QV=0$, namely \eqref{1st} - \eqref{7th} are satisfied identically as we indicate in appendix \ref{sec:check}. The equations \eqref{D_Gmn1} - \eqref{cons_theta=0} are the equations which are needed to do the $\theta$ expansion of the superfields completely which is done in the next section.

\sectiono{$\theta$ Expansion} \label{thetaex}
As mentioned in the previous sections, the lowest components of the superfields $\Psi_{s\alpha},B_{mnp}$ and $G_{mn}$ are the physical fields $\psi_{s\alpha},b_{mnp}$ and $g_{mn}$ respectively. The higher $\theta$ components of the superfields contain the same physical fields in a more involved manner. In this section, we shall outline the procedure to determine the $\theta$ expansion of these superfields in terms of the physical fields exclusively. We recall the key equations from previous section which we shall need below
\be
D_{\alpha}\Psi_{s\beta}&=& \f{1}{16} G_{sm}\gamma^{m}_{\alpha\beta}+\f{i}{24}k_mB_{nps}(\gamma^{mnp})_{\alpha\beta}-\f{i}{144}k^mB^{npq}(\gamma_{smnpq})_{\alpha\beta} \label{D_Psi}\\ [.19in]
D_{\alpha}B_{mnp}&=&12(\gamma_{[mn}\Psi_{p]})_{\alpha}+24\alpha'k^t k_{[m}(\gamma_{|t|n}\Psi_{p]})_{\alpha}   \\[.19in]
D_{\alpha}G_{sm}&=&16i k^{p}(\gamma_{p(s}\Psi_{m)})_\alpha \label{D_Gmn}
\ee
These equations are sufficient to obtain the $\theta$ expansion of all the superfields to all orders in $\theta$ once we specify the 
$\theta$ independent components of $B_{mnp},G_{sm}$ and $\Psi_{s\alpha}$. 
 Intuitively, we can see how 
these equations will determine the higher $\theta$ components once the lowest components are specified - if we equate the $\theta^{\ell}$ components on both sides of these equations, we shall have 
$\theta^{\ell}$ components of the superfields on the right hand side. But on the left hand side, because we have a covariant derivative $D_{\alpha}$, we shall always have $\theta^{\ell+1}$ and $\theta^{\ell-1}$
component of the superfield on which $D_{\alpha}$ acts. Thus, the higher components can be determined in terms of the lower components.

\vspace*{.06in}We denote the $\theta$ expansion of the superfields as
\ben 
\Psi_{s\beta}&=& \sum_{n=0}^{16}\psi_{s\beta\alpha_{1}\alpha_{2}\cdots\alpha_{n}}\theta^{\alpha_1}\theta^{\alpha_2}\cdots\theta^{\alpha_n} \label{Psi_expan}\\
B_{mnp}&=& \sum_{n=0}^{16}b_{mnp\alpha_{1}\alpha_{2}\cdots\alpha_{n}}\theta^{\alpha_1}\theta^{\alpha_2}\cdots\theta^{\alpha_n}         \\
G_{mn}&=&\sum_{n=0}^{16}g_{mn\alpha_{1}\alpha_{2}\cdots\alpha_{n}}\theta^{\alpha_1}\theta^{\alpha_2}\cdots\theta^{\alpha_n} \label{Gmn_expan}
\een
For each of
 the superfields, the fermionic and bosonic degrees of freedom occur either at even or odd $\theta$ components only. For example in $\Psi_{s\beta}$, the fermionic field $\psi_{s\beta}$ appears at even 
 $\theta$ components and the bosonic fields $g_{mn}$ and $b_{mnp}$ appear at odd $\theta$ components respectively. In the case of $B_{mnp}$ and 
 $G_{mn}$, the bosonic fields appear at the even $\theta$ components and the fermionic field appear at odd $\theta$ components. While moving the fermionic objects such as $\theta^\alpha$ across these various components, it is helpful to keep in mind the aforementioned points. 

\vspace*{.06in}On substituting \eqref{Psi_expan}-\eqref{Gmn_expan} into \eqref{D_Psi}-\eqref{D_Gmn} and comparing the $(\ell- 1)^{th}$ component on both sides, we find 
\ben
&&\hspace*{-.5in}(-1)^{\ell-1}\Bigl[\ell\;\psi_{s\beta\alpha_1\alpha_2\cdots\alpha_{\ell}}+ik_r(\gamma^r)_{\alpha_1\alpha_2}\psi_{s\beta\alpha_3\cdots\alpha_\ell}
\Bigl]\non\\
&=&\left[\f{1}{16}(\gamma^m)_{\beta\alpha_1}\ g_{sm\alpha_2\cdots\alpha_\ell}-\f{i}{24}k_m b_{nps\alpha_2\cdots\alpha_\ell}(\gamma^{mnp})_{\beta\alpha_1}
-\f{i}{144}k_mb_{npq\alpha_2\cdots\alpha_\ell}(\gamma_{s}^{\;\;mnpq})_{\beta\alpha_1}\right]   						\label{D_Psi_comp}\\[.35in]
&&\hspace*{-.5in}\ell\;b_{mnp\alpha_1\alpha_2\cdots\alpha_{\ell}}+ik_r(\gamma^r)_{\alpha_1\alpha_2}b_{mnp\alpha_3\cdots\alpha_\ell}
\non\\
&&\hspace*{-.25in}=(-1)^{\ell}\biggl[12(\gamma_{[mn})_{\alpha_1}^{\;\;\sigma}\psi_{p]\sigma\alpha_2\cdots\alpha_\ell}
+24\alpha'k^t k_{[m}(\gamma_{|t|n})_{\alpha_1}^{\;\;\;\sigma}\psi_{p]\sigma \alpha_2\cdots\alpha_\ell}\biggl] 			\label{D_Bmnp_comp}\\[.35in]
&&\hspace*{-.4in}\ell\; g_{mn\alpha_1\alpha_2\cdots\alpha_{\ell}}+ik_r(\gamma^r)_{\alpha_1\alpha_2}g_{mn\alpha_3\cdots\alpha_\ell}
= (-1)^{\ell}16\, i\, k^p(\gamma_{p(s})_{\alpha_1}^{\;\;\sigma}\psi_{m)\sigma\alpha_2\cdots\alpha_\ell}				\label{D_Gmn_comp}
\een
where, $(\ell-1)=1,2,\cdots ,16$.

\vspace*{.06in}The higher $\theta$ components of the superfields can now be fixed using the above equations recursively. We start with the $\ell=1$ component of \eqref{D_Psi_comp} which fixes the $O(\theta)$ component of $\Psi_{s\beta}$ to be
\ben
\psi_{s\beta\alpha_1}\theta^{\alpha_1}=
\f{1}{16}(\gamma^m\theta)_{\beta}\ g_{sm}-\f{i}{24}k_m b_{nps}(\gamma^{mnp}\theta)_{\beta}
-\f{i}{144}k_mb_{npq}(\gamma_{s}^{\;\;mnpq}\theta)_{\beta}
\label{ell=1psi}								
\een
Next, using the fact that 1-form $\gamma^m_{\alpha\beta}$ is symmetric in its spinor indices, the equation \eqref{D_Psi_comp} gives for $\ell=2$
\ben
\psi_{s\beta\alpha_1\alpha_2}\theta^{\alpha_1}\theta^{\alpha_2}=
\f{1}{16}(\gamma^m\theta)_{\beta}\ g_{sm\alpha_2}\theta^{\alpha_2}+\f{i}{24}k_m(\gamma^{mnp}\theta)_{\beta} b_{nps\alpha_2}\theta^{\alpha_2}
+\f{i}{144}k_m(\gamma_{s}^{\;\;mnpq}\theta)_{\beta}b_{npq\alpha_2}\theta^{\alpha_2} \non\\	
\label{ell=2psi}							
\een
where, we have used the fact that $b_{mnp\alpha}$ and $g_{mn\alpha}$ are Grassmann odd. They can be determined using the $\ell=1$ components of equations \eqref{D_Bmnp_comp} and \eqref{D_Gmn_comp} respectively as
\ben
b_{mnp\alpha_2}&=&-12(\gamma_{[mn})_{\alpha_2}^{\;\;\sigma}\psi_{p]\sigma}
-24\alpha'k^t k_{[m}(\gamma_{|t|n})_{\alpha_2}^{\;\;\;\sigma}\psi_{p]\sigma }\non\\[.15in]						
g_{mn\alpha_2}
&=&-16\, i\, k^p(\gamma_{p(m})_{\alpha_2}^{\;\;\sigma}\psi_{n)\sigma}									
\een
These can be substituted in equation \eqref{ell=2psi} to fix the $O(\theta^2)$ component of $\Psi_{s\beta}$ completely. 

\vspace*{.06in}At the next oder in $\theta$, the $\psi_{s\beta\alpha_1\alpha_2\alpha_3}$ will require $g_{mn\alpha_2\alpha_3}$ and $b_{mnp\alpha_2\alpha_3}$. These can again be determined using equations \eqref{D_Bmnp_comp} and \eqref{D_Gmn_comp} along with order $\theta$ result \eqref{ell=1psi}.

\vspace*{.06in}Executing this process recursively determines all the $\theta$ components of the superfields in terms of the physical fields.  Equation \eqref{D_Psi_comp} relates the $\ell^{th}$ component of $\Psi_{s\beta}$ to the $(\ell-2)^{th}$ component of $\Psi_{s\beta}$ and $(\ell-1)^{th}$ components of $B_{mnp}$ and
 $G_{sm}$.  Equation \eqref{D_Bmnp_comp} relates the $\ell^{th}$ component of $B_{mnp}$ to the $(\ell-2)^{th}$ component of $B_{mnp}$ and $(\ell-1)^{th}$ component of $\Psi_{s\beta}$. Finally, the equation \eqref{D_Gmn_comp} determines $\ell^{th}$ component of $G_{mn}$ in terms of $(\ell-2)^{th}$ component of $G_{mn}$ and $(\ell-1)^{th}$ component of $\Psi_{s\alpha}$. These statements can be schematically represented as 
 \ben 
 D^{(\ell+1)}\Psi_{s\beta}&\sim &D^\ell G_{sm}+D^\ell B_{mnp}\non\\
 D^\ell B_{mnp}&\sim&D^{(\ell-1)}\Psi_{s\beta} \non\\
 D^\ell G_{mn}&\sim & D^{(\ell-1)}\Psi_{s\beta} \non
 \een
 where for any superfield $V$, $D V\equiv (\theta^{\alpha}D_{\alpha} V )\big|_{\theta=0}$, so that $D^\ell$ picks up the $\ell^{th}$ $\theta$ component of the superfield it acts on.

\vspace*{.06in}We shall now give the explicit expressions for the superfields $\Psi_{s\beta}$ and $B_{mnp}$ upto order $\theta^3$. The higher order components are tedious but straightforward to evaluate if required. Using the procedure described above, the $\theta$ expansion of the $\Psi_{s\beta}$ field is given by
\be
\Psi_{s\beta}&=&\psi_{s\beta}+\f{1}{16}(\gamma^m\theta)_\beta\ g_{sm}-\f{i}{24}(\gamma^{mnp}\theta)_\beta k_{m}b_{nps}-\f{i}{144}(\gamma_{s}^{\;\;npqr}\theta)_{\beta}k_{n}b_{pqr} \non\\
&&-\f{i}{2}k^p(\gamma^m\theta)_{\beta}(\psi_{(m}\gamma_{s)p}\theta)-\f{i}{4}k_{m}(\gamma^{mnp}\theta)_{\beta}(\psi_{[s}\gamma_{np]}\theta)-\f{i}{24}(\gamma_{s}^{\;\;mnpq}\theta)_{\beta}k_{m}(\psi_{q}\gamma_{np}\theta)\non\\
&&-\f{i}{6}\alpha'k_mk^rk_s(\gamma^{mnp}\theta)_{\beta}
(\psi_p\gamma_{rn}\theta)+{\f{i}{288}}\alpha'(\gamma^{mnp}\theta)_\beta k_mk^rk_s(\theta\gamma^q_{\;\;nr}\theta)\ g_{pq}\non\\
&&-{\f{i}{192}}(\gamma^{mnp}\theta)_\beta k_m(\theta\gamma^q_{\;\;[np}\theta) g_{s]q}-\f{i}{1152}(\gamma_{smnpq}\theta)_{\beta}k^{m}(\theta\gamma_{npt}\theta)\ g^{qt}\non\\
&&-\f{i}{96} k^p(\gamma^m\theta)_{\beta}(\theta\gamma_{pq(s}\theta)\ g_{m)q}-\f{1}{1728}(\gamma^{mnp}\theta)_\beta k_m(\theta\gamma^{tuvw}_{\;\;\;\;\;\;\;\;\;nps}\theta)k_tb_{uvw}\non\\
&&-{\f{1}{864\alpha'}}(\gamma_s\theta)_{\beta}(\theta\gamma^{npq}\theta)b_{npq}-\f{1}{10368}(\gamma_{s}^{\;\;mnpq}\theta)_{\beta}k_{m}(\theta\gamma_{tuvwnpq}\theta)k^tb^{uvw}\non\\
&&-{\f{1}{864}}(\gamma^m\theta)_{\beta}(\theta\gamma^{npq}\theta)b_{npq}k_{m}k_s-\f{1}{576}(\gamma_{smnpq}\theta)_{\beta}k^{m}(\theta\gamma^{tun}\theta)b_u^{\;\;pq}k_t\non\\
&&-\f{1}{96\alpha'}(\gamma^m\theta)_{\beta}(\theta\gamma^{qr}_{\;\;\;\;(s}\theta)b_{m)rq}+\f{1}{96}(\gamma^m\theta)_{\beta}(\theta\gamma^{nqr}\theta)k_nk_{(s}b_{m)qr}\non\\
&&+{\f{1}{96}}(\gamma^{mnp}\theta)_\beta k_m(\theta\gamma^r_{\;\;q[n}\theta)b_{ps]r} k^{q}\;\; +\;\;O(\theta^4)
\ee
Similarly, the $\theta$ expansion of the superfield $B_{\alpha\beta}$ is given by
\be
B_{\alpha\beta}&=&\gamma^{mnp}_{\alpha\beta}\Biggl[b_{mnp}+12(\psi_p\gamma_{mn}\theta)+24\alpha'k^rk_{m}(\psi_p\gamma_{rn}\theta)+\f{3}{8}(\theta\gamma_{mn}^{\;\;\;\;\;q}\theta)\ g_{pq}-\f{3i}{4}(\theta\gamma^{tu}_{\;\;\;\;m}\theta)k_{t}b_{unp}\non\\
&&\hspace*{.4in}+\f{3}{4}\alpha'k^rk_{m}(\theta\gamma_{rn}^{\;\;\;\;\;q}\theta)\ g_{pq}
-\f{i}{24}(\theta\gamma_{tuvwmnp}\theta)k^{t}b^{uvw}-\frac{1}{6} i k_{{s}} \left(\psi _{{v}} \gamma _{{t} {u}} \theta \right) \left(\theta  \gamma _{stuv m n p} \theta \right)\non\\
&&\hspace*{.4in}-4 i \alpha  k_{{s}} k_{{t}} k_m \left(\theta  \gamma _{{t} {u} n} \theta \right) \left(\psi _p \gamma _{{s} {u}} \theta \right)+i k_{{s}} \left(\theta  \gamma _{{t} m n} \theta \right) \left(\psi _p \gamma _{{s} {t}} \theta \right)+i k_{{s}} \left(\theta  \gamma _{{t} m n} \theta \right) \left(\psi _{{t}} \gamma _{{s} p} \theta \right)\non\\
&&\hspace*{.4in}+2 i k_{{s}} \left(\theta  \gamma _{{s} {t} m} \theta \right) \left(\psi _n \gamma _{{t} p} \theta \right)-i k_{{s}} \left(\theta  \gamma _{{s} {t} m} \theta \right) \left(\psi _{{t}} \gamma _{n p} \theta\right)\;+\; O(\theta^4)\Biggl]
\ee
Since all the superfields appearing in the first massive vertex operator can be expressed fully using the superfields $B_{\alpha\beta}$ and $\Psi_{m\alpha}$, the above results are enough to write down the $\theta$ expansion of the unintegrated vertex operator upto $O(\theta^3)$. 
In the next section, we shall give this result for the full vertex operator of the $b_{mnp}$ field. 

\vspace*{0.06in}For completeness, we also give the $\theta$ expansion of $G_{sm}$ upto $O(\theta^3)$
\be
G_{sm}&=&g_{sm}-16i k^p(\psi_{(m}\gamma_{s)p}\theta)+\f{i}{2}k^p(\theta\gamma_{p(m}\gamma^n\theta)\ g_{s)n}+\f{1}{3}k^p(\theta\gamma_{p(m}\gamma^{tqr}\theta) k_{|t}b_{qr|s)} \non\\
&&+\f{1}{18}k^p(\theta\gamma_{p(m}\gamma_{s)}^{\;\;ntqr}\theta)k_{n}b_{tqr}+\f{8}{9}\alpha'k_t k^pk^rk_{(s}(\theta\gamma_{m)p}\gamma^{tnq}\theta)
(\psi_{q}\gamma_{rn}\theta)\non\\
&&-\f{8}{3}k^t k^p(\theta\gamma_{p(m}\gamma^n\theta)(\psi_{(n}\gamma_{s))t}\theta)-\f{4}{3}k_{t} k^p(\theta\gamma_{p(m}\gamma^{tnq}\theta)(\psi_{[s)}\gamma_{nq]}\theta)\non\\
&&-\f{2}{9}k_{t} k^p(\theta\gamma_{p(m}\gamma_{s)}^{\;\;tnrq}\theta)(\psi_{q}\gamma_{nr}\theta)\;+\; O(\theta^4)\non
\ee

\sectiono{3-point tree $\big\langle AAb \big\rangle$ amplitude} \label{amplical} 
One of the applications of the results of previous section is in computing scattering amplitudes involving the massive states in pure spinor formalism. Just for illustration, in this section, we consider the 3-point tree amplitude involving 2 gluon fields\footnote{The $\theta$ expansion of the massless fields is given in appendix \ref{sec:massless}.} (denoted by $ a_m^{(i)}$) and the 3-form field $b_{mnp}$. This amplitude was also considered in \cite{Park}. However, our result for $\theta$ expansion is significantly different from that of \cite{Park}. Hence, we compute the contribution of terms in the massive vertex operator upto $O(\theta^3)$ to this amplitude and check that our result agrees with the corresponding kinematic factor in the RNS formalism. The full amplitude acquires also the contribution from higher $\theta$ components which we do not consider in this paper (see conclusion for further comments regarding the full amplitude). 

\vspace*{.06in}Since we shall only compute the 3-point function on the disk, the equation \eqref{treeamp} tells us that we need only the unintegrated vertex operator
to compute the amplitude  
\be
\mathcal{A}_3=\langle V^1V^2V^3 \rangle 
\ee
where, $V^i$ are the unintegrated vertex operators of interest (massive or massless). 

\vspace*{.06in}The pure spinor measure is defined such that the bracket $\langle ... \rangle$ gives non-zero answer if and only if there are three $\lambda$ and five $\theta$ zero mode inside it. Symbolically, this is often abbreviated as $\langle \lambda^3 \theta^5 \rangle \sim 1$. More precisely, the pure spinor measure is normalized as 
\be 
\langle (\lambda \gamma^m \theta)(\lambda \gamma^n \theta)(\lambda \gamma^p \theta)(\theta \gamma_{m n p} \theta) \rangle = 1
\ee
We now outline the procedure for computing the tree amplitudes. Given 3 external states whose tree level scattering we wish to compute, the basic strategy is as follows :
\begin{itemize}
\item Write down the $\theta$ expansion of each vertex operator $V^i$ to the desired order.

\item Total number of $\theta$ in the product $V^1 V^2 V^3$ for non-zero contribution to the amplitude must be exactly equal to 5. So, from the product, we keep only those terms which have exactly five factors of $\theta$. 

\item Since each unintegrated vertex operator $V^i$ has ghost number 1, they come with a single factor of $\lambda^\alpha$. Therefore each term in the product $V^1 V^2 V^3$ always has exactly three factors of $\lambda^\alpha$.

\item Express every physical field in terms of its polarization and plane wave basis. For example, 
\be 
 A^{m_1...m_n}_{\alpha_1...\alpha_k}(X)= a ^{m_1...m_n}_{\alpha_1...\alpha_k} \, e^{ik\cdot X}\non
\ee
where $a ^{m_1...m_n}_{\alpha_1...\alpha_k}$ is the constant tensor-spinor polarization corresponding to the physical field $A^{m_1...m_n}_{\alpha_1...\alpha_k}(X)$.

\item Compute the correlation function  $\langle :e^{i k_1 \cdot X}::e^{i k_2 \cdot X}::e^{i k_3 \cdot X} f(X^m):\rangle_{Disk}$ separately, where typically 
$f(X^m)$ is just a product of various $\partial X^m$.

\item The only thing left to compute at this stage is the correlation function in Pure Spinor Superspace. The relevant correlation functions were computed in \cite{Berkovits_Mafra}. For completeness, we give a list of correlators, used in this paper, in the appendix \ref{PureSuperspace}.

\end{itemize}
All the computations, even though straightforward in principle, are quite tedious to do by hand (especially since the $\theta$ expansions of massive vertex operators for all the external states contain a lot of terms at higher order). The use of Mathematica package Gamma \cite{Gamma} was indispensable in these calculations.

\vspace*{.06in}To compute the amplitude, we need to ``multiply'' the vertex operator of $b_{mnp}$ field with the vertex operators of gluons and pick up the terms which have exactly five $\theta$.
\begin{table}
\begin{center}
\vspace{0.15in}
\setlength{\arrayrulewidth}{.25mm}
\renewcommand{\arraystretch}{1.5}
\begin{tabular}{ | p{1.6cm} | p{1.6cm}| p{1.6cm}|} 
\hline
{ $V_a^{(1)}$} 
& 
{$V_a^{(2)}$}
&
{$V_b$}
\\ 
\hline
{ 1} 
&
{1 }
&
{3}
\\ 
\hline
{1} 
&
{3}
&
{1}
\\
\hline
{3} 
& 
{1}
&
{1}
 \\ 
\hline
\end{tabular}
\end{center}
\caption{Possible distribution of five $\theta$ in the vertex operators due to first three terms of \eqref{massive_bmnp}.
\label{t1}
} 
\vskip -.1in
\end{table}
Now, the on-shell vertex operator for the first massive states is given by 
\be
V=\partial \theta^\beta \lambda^\alpha B_{\alpha\beta}(X,\theta) +\Pi^m\lambda^\alpha H_{m\alpha}(X,\theta)+N^{mn}\lambda^\alpha F_{\alpha mn}(X,\theta)+d_\beta\lambda^\alpha C^\beta_{\;\alpha}(X,\theta)\label{massive_bmnp}
\ee
where the on-shell expression of superfields appearing here are given in equation \eqref{2.14}.

\vspace*{.06in}The $d_\alpha$ has a non trivial OPE with $\theta^\alpha$ and can reduce the number of zero modes of $\theta$ by one in the correlator. The massless vertex operator contains no $d_\alpha$ term and its $\theta$ expansion starts at $O(\theta)$. Since the $d_\alpha$ term of the massive vertex operator will reduce the number of $\theta$ zero modes from the massless vertices by one at a time, the minimum number of $\theta$ zero modes supplied by the two massless vertices can be one for the amplitude. This means that the contribution of the $d_\beta\lambda^\alpha C^\beta_{\;\alpha}$ term in the massive vertex operator will require the knowledge of $\theta$ expansion upto $O(\theta^4)$. Thus, we focus here only on the contributions coming from the first 3 terms in the right hand side of equation \eqref{massive_bmnp}.  The possible $\theta$ distribution for these terms is shown in table \ref{t1}. From this, it is clear that these three terms require the $\theta$ expansion only upto $O(\theta^3)$. 

\vspace*{.06in}Using equations \eqref{2.14}, \eqref{3.21} and the result of $\theta$ expansion for the superfields $B_{mnp}$ and $\Psi_{m\alpha}$ given in previous section, the $\theta$ expansion for the four terms in the above vertex operator for $b_{mnp}$ field upto $O(\theta^3)$ is given by
\be
\partial \theta^\beta \lambda^\alpha B_{\alpha\beta}(X,\theta)
&=&\biggl[(\lambda\gamma^{mnp}\p\theta)b_{mnp}
-\f{i}{24}(\lambda\gamma^{mnp}\p\theta)(\theta\gamma_{tuvwmnp}\theta)k^{t}b^{uvw}\non\\
&&-\f{3i}{4}(\lambda\gamma^{mnp}\p\theta)(\theta\gamma^{tu}_{\;\;\;\;m}\theta)k_{t}b_{unp}\biggl]\non
\ee
\be
d_\beta\lambda^\alpha C^\beta_{\;\alpha}(X,\theta)
&=&\biggl[\f{i}{2}(d\gamma^{smnp}\lambda)k_{s}b_{mnp}+\f{1}{48}(d\gamma^{smnp}\lambda)(\theta\gamma_{tuvwmnp}\theta)k^{t}k_{s}b^{uvw}\non\\
&&+\f{3}{8}(d\gamma^{smnp}\lambda)(\theta\gamma^{tu}_{\;\;\;\;m}\theta)k_{t}k_{s}b_{unp}\biggl]\non
\ee
\be
&&\hspace*{-.25in}\Pi^m\lambda^\alpha H_{m\alpha}(X,\theta)\non\\
&=& \biggl[3i\Pi^s(\lambda\gamma^{mnp}\theta) k_{m}b_{nps}+\f{i}{2}\Pi^s(\lambda\gamma_{s}^{\;\;npqr}\theta)k_{n}b_{pqr}-\f{3}{4}\Pi^s(\lambda\gamma^m\theta)(\theta\gamma^{nqr}\theta)k_nk_{(s}b_{m)qr} \non\\
&&+\f{3}{4\alpha'}\Pi^s(\lambda\gamma^m\theta)(\theta\gamma^{qr}_{\;\;\;\;(s}\theta)b_{m)rq}+\f{1}{12\alpha'}\Pi^s(\lambda\gamma_s\theta)(\theta\gamma^{npq}\theta)b_{npq}+\f{1}{12}\Pi^s(\lambda\gamma^m\theta)(\theta\gamma^{npq}\theta)b_{npq}k_{m}k_s\non\\
&&-\f{3}{4}\Pi^s(\lambda\gamma^{mnp}\theta) k_m(\theta\gamma^r_{\;\;q[n}\theta)b_{ps]r} k^{q}+\f{1}{24}\Pi^s(\lambda\gamma^{mnp}\theta) k_m(\theta\gamma^{tuvw}_{\;\;\;\;\;\;\;\;\;nps}\theta)k_tb_{uvw} \non\\
&&+\f{1}{8}\Pi^s(\lambda\gamma_{smnpq}\theta)k^{m}(\theta\gamma^{tun}\theta)b_u^{\;\;pq}k_t+\f{1}{144}\Pi^s(\lambda\gamma_{s}^{\;\;mnpq}\theta)k_{m}(\theta\gamma_{tuvwnpq}\theta)k^tb^{uvw}\biggl]\non
\ee
\be
&&\hspace*{-.3in}N^{xs}\lambda^\alpha F_{\alpha xs}(X,\theta)\non\\
&=&-\f{1}{1152\alpha'}N^{xs}\biggl[-864 (\lambda \gamma ^{{n}}\theta)  b_{{n} x s}-216 (\lambda \gamma ^{{n} {p}}_{\;\;\;\; {s}}\theta)  b_{{n} {p} {x}} -3240 \alpha  (\lambda \gamma ^{{m} {n} {r}}\theta)  b_{{m} {n} x} k_{s} k_{{r}}\non\\
&&-72 (\lambda \gamma _{{n} {p} {q} {x} {s}}\theta)  b^{{n} {p} {q}} -648 \alpha  (\lambda \gamma _{{m} {n} {p} {x} {t}}\theta)  b^{{m} {n} {p}}  k_{{s}} k^{{t}} -378 i (\lambda \gamma _{{r}}\theta) (\theta\gamma^{{m} {n} {r}}\theta) b_{{m} {n} x} k_{s}\non\\
&& -72 i (\lambda \gamma _{{t}}\theta) (\theta\gamma^{{n} {r} {t}}\theta) b_{{n} x s} k_{{r}}-378 i \alpha  (\lambda \gamma ^{{t}}\theta) (\theta\gamma^{{m} {n} {r}}\theta) b_{{m} {n} x} k_{s} k_{{r}} k_{{t}}-378 i (\lambda \gamma _{{p}}\theta) (\theta\gamma_{{m} {n} {x}}\theta) b^{{m} {n} {p}}  k_{{s}}\non\\
&&+84 i (\lambda \gamma _{{x}}\theta) (\theta\gamma^{{n} {p} {q}}\theta) b_{{n} {p} {q}}  k_{{s}}+144 i (\lambda \gamma ^{{p}}\theta) (\theta\gamma^{{n}}_{\;\;\; {s} {t}}\theta) b_{{n} {p} {x}}  k^{{t}}-12 i (\lambda \gamma ^{{v}}\theta) (\theta\gamma_{{n} {p} {q} {x} {s} {u} {v}}\theta) b^{{n} {p} {q}}  k^{{u}} \non\\
&&+54 i (\lambda \gamma _{s {r} {t}}\theta) (\theta\gamma^{{m} {n} {t}}\theta) b_{{m} {n} x} k^{{r}}+3 i (\lambda \gamma _{{s}}^{\;\;\; {v} {w}}\theta) (\theta\gamma_{{n} {p} {q} {x} {u} {v} {w}}\theta) b^{{n} {p} {q}}  k^{{u}}+18 i (\lambda \gamma _{{p} {q} {s}}\theta) (\theta\gamma_{{n} {x} {u}}\theta) b^{{n} {p} {q}}  k^{{u}} \non\\
&&-36 i (\lambda \gamma ^{{p}}_{\;\;\; {s} {u}}\theta) (\theta\gamma^{{n} {t} {u}}\theta) b_{{n} {p} {x}}  k_{{t}}+54 i (\lambda \gamma _{{p} {s} {t}}\theta) (\theta\gamma_{{m} {n} {x}}\theta) b^{{m} {n} {p}}  k^{{t}}-12 i (\lambda \gamma _{{x} {s} {t}}\theta) (\theta\gamma_{{m} {n} {p}}\theta) b^{{m} {n} {p}}  k^{{t}}\non\\
&&-540 i \alpha  (\lambda \gamma ^{{n} {t} {u}}\theta) (\theta\gamma^{{m} {r}}_{\;\;\;\;\; {u}}\theta) b_{{m} {n} x} k_{s} k_{{r}} k_{{t}}+45 i \alpha  (\lambda \gamma ^{{v} {w}}_{\;\;\;\; \;{u} }\theta) (\theta\gamma_{{m} {n} {p} {x} {t} {v} {w}}\theta) b^{{m} {n} {p}}  k_{{s}} k^{{t}} k^{{u}}\non\\
&&+270 i \alpha  (\lambda \gamma _{{n} {p} {u}}\theta) (\theta\gamma_{{m} {x} {t}}\theta) b^{{m} {n} {p}}  k_{{s}} k^{{t}} k^{{u}} -54 i \alpha  (\lambda \gamma _{{p} {x} {u}}\theta) (\theta\gamma_{{m} {n} {t}}\theta) b^{{m} {n} {p}}  k_{{s}} k^{{t}} k^{{u}} \non\\
&&-18 i (\lambda \gamma _{{p} {q} {x} {s} {v}}\theta) (\theta\gamma_{{n}}^{\;\;\; {u} {v}}\theta) b^{{n} {p} {q}}  k_{{u}}-9 i \alpha  (\lambda \gamma _{{x} {u}}^{\;\;\;\;\; {v} {w} {q}}\theta) (\theta\gamma_{{m} {n} {p} {t} {v} {w} {q}}\theta) b^{{m} {n} {p}}  k_{{s}} k^{{t}} k^{{u}}\non\\
&&-i (\lambda \gamma _{{x} {s} {v} {w} {m}}\theta) (\theta\gamma_{{n} {p} {q} {u}}^{\;\;\;\;\;\;\;\; {v} {w} {m}}\theta) b^{{n} {p} {q}}  k^{{u}}-162 i \alpha  (\lambda \gamma _{{n} {p} {x} {u} {v}}\theta) (\theta\gamma_{{m} {t}}
^{\;\;\;\; {v}}\theta) b^{{m} {n} {p}}  k_{{s}} k^{{t}} k^{{u}} \biggl]\non
\ee

We now consider the above mentioned terms in the massive vertex operator and evaluate their contribution. We shall indicate some steps for few terms below. We put the two gluons at $z_1$ and $z_2$ and $b_{mnp}$ field at $z_3$ on the world-sheet and use the notation 
\be
a_m^{(1)}(X) =e_m^{(1)} e^{ip_1\cdot X}\qquad,\qquad a_m^{(2)}(X) =e_m^{(2)} e^{ip_2\cdot X}\qquad,\qquad b_{mnp}=e_{mnp} e^{ik\cdot X}\non
\ee
where, the polarization tensors satisfy the transversality conditions
\be
e_m^{(1)}p_1^m=0\qquad,\quad e_m^{(2)}p^m_2=0\qquad,\qquad e_{mnp}k^m=0\label{trans_condi}
\ee
For the first type of terms, $\partial \theta^\beta \lambda^\alpha B_{\alpha\beta}$ in the $b_{mnp}$ vertex operator, we note that it contains either zero or two $\theta$s. Hence, it will never contribute to this amplitude. So, we start with the terms of the type $\Pi^m\lambda^\alpha H_{m\alpha}$. Using the expression of massless vertex operator given in appendix \ref{sec:massless}, the first term of $\Pi^m\lambda^\alpha H_{m\alpha}$ gives (noting that only the term containing $\p X^m$ in $\Pi^m$ will be relevant)
\be
I&=&\left\langle\left(\f{1}{2}a_r(\lambda\gamma^r\theta)
 \right)\left(
-\f{1}{32}F_{tu}(\lambda\gamma_v\theta)(\theta\gamma^{tuv}\theta) \right) 3i\Pi^s(\lambda\gamma^{mnp}\theta) k_{m}b_{nps} \right\rangle\non\\
&&+\left\langle\left(
-\f{1}{32}F_{tu}(\lambda\gamma_v\theta)(\theta\gamma^{tuv}\theta) \right)\left(\f{1}{2}a_r(\lambda\gamma^r\theta)
\right) 3i\Pi^s(\lambda\gamma^{mnp}\theta) k_{m}b_{nps} \right\rangle\non\\
&=&-\f{3i}{64}k_{m}\left(e_r^{(1)}f_{tu}^{(2)}-e_r^{(2)}f_{tu}^{(1)}\right)e_{nps}\left\langle (\lambda\gamma^r\theta)
(\lambda\gamma_v\theta)(\lambda\gamma^{mnp}\theta)(\theta\gamma^{tuv}\theta) \right\rangle\Gamma^s\non\\
&=&\f{1}{1920\alpha'}e^{mnp}e_p^{(1)}e_n^{(2)}\Gamma_m\label{pi_first}
\ee
where, $f_{mn}=ip_me_n-ip_ne_m$ and $\Gamma^m$ is the world-sheet correlator involving the $X$ fields
\be
\Gamma^m(z_1,z_2,z_3)\equiv\left\langle :e^{ip_1\cdot X(z_1)}::e^{ip_2\cdot X(z_2)}::e^{ik\cdot X(z_3)}\p X^m(z_3): \right\rangle=i\alpha'\left(\f{p_1^m z_{23}+ p^m_2 z_{13}}{z_{12} }\right)\label{gamma_q}
\ee
where, $z_{ij}=z_{i}-z_{j}$.

\vspace*{.06in}In going to the last line of \eqref{pi_first}, we have used the pure spinor identity \eqref{C.2} given in appendix \ref{PureSuperspace}.

\vspace*{.06in}Now, it might appear that this correlator depends upon the worldsheet coordinates $z_i$ through $\Gamma^m$. However, the momentum conservation and the transversality of the polarization tensors ensure that this is not the case. To see this, we simplify the last line of \eqref{pi_first} using \eqref{gamma_q} as follows
\be
I&=&\f{i}{1920}e^{mnp}e_p^{(1)}e_n^{(2)}\left(\f{(p_1)_m z_{23}+ (p_2)_m z_{13}}{z_{12} }\right)\non\\
&=&\f{i}{1920}e^{mnp}e_p^{(1)}e_n^{(2)}(p_2)_m\left(\f{- z_{23}+  z_{13}}{z_{12} }\right)\non\\
&=&\f{i}{1920}e^{mnp}e_p^{(1)}e_n^{(2)}(p_2)_m\non
\ee
In going to the second line, we have used the momentum conservation $p_1+p_2+k=0$ and the transversality condition $e^{mnp}k_m=0$.

\vspace*{.06in}In an identical manner, the contribution of the rest of the terms in $\Pi^m\lambda^\alpha H_{m\alpha}$ can be evaluated. One gets the same tensor structure from these terms and their total contribution is given by
\be
\f{31i}{13440}e^{mnp}e_{p}^{(1)}e_{n}^{(2)}(p_2)_m\label{1st_contri}
\ee

Next, we consider the $N^{mn}\lambda^\alpha F_{\alpha mn}$ term of the massive vertex operator. Its first term gives,
\be
I'&\equiv&\left\langle\left(\f{1}{2}a_r^{(1)}(\lambda\gamma^r\theta)
 \right)\left(
-\f{1}{32}F_{tu}^{(2)}(\lambda\gamma_v\theta)(\theta\gamma^{tuv}\theta) \right) \f{864}{1152\alpha'}N^{xs}(\lambda\gamma^n\theta)b_{nxs} \right\rangle\non\\
&&+\left\langle\left(
-\f{1}{32}F_{tu}^{(1)}(\lambda\gamma_v\theta)(\theta\gamma^{tuv}\theta) \right)\left(\f{1}{2}a_r^{(2)}(\lambda\gamma^r\theta)
\right) \f{864}{1152\alpha'}N^{xs}(\lambda\gamma^n\theta)b_{nxs} \right\rangle\non\\
&=&-\f{27}{2304\alpha'}e_r^{(1)}f_{tu}^{(2)}e_{nxs}\left\langle (\lambda\gamma^r\theta)
(\lambda\gamma_v\theta)(\theta\gamma^{tuv}\theta)N^{xs}(\lambda\gamma^n\theta)\right\rangle\Gamma(z_1,z_2,z_3)\non\\
&&-\f{27}{2304\alpha'}e_r^{(2)}f_{tu}^{(1)}e_{nxs}\left\langle
(\lambda\gamma_v\theta) (\lambda\gamma^r\theta)(\theta\gamma^{tuv}\theta)
N^{xs}(\lambda\gamma^n\theta)\right\rangle\Gamma(z_1,z_2,z_3)\label{5.8}
\ee
where, $\Gamma$ is the world-sheet correlator
\be
\Gamma(z_1,z_2,z_3)&\equiv&\left\langle :e^{ip_1\cdot X(z_1)}::e^{ip_2\cdot X(z_2)}::e^{ik\cdot X(z_3)}: \right\rangle\;=\;\f{z_{23}z_{13}}{z_{12}}\non
\ee
To evaluate the pure spinor correlators, we first need to eliminate the $N^{xs}$ field by using its OPE with $\lambda^\alpha$. We illustrate it with the second correlator of \eqref{5.8}
\be 
&&\left\langle:(\lambda\gamma_v\theta):(z_1) :(\lambda\gamma^r\theta)(\theta\gamma^{tuv}\theta):(z_2) :N^{xs}(\lambda\gamma^n\theta):(z_3)\right\rangle\non\\
&=&\oint_{z_3}\f{dw}{w-z_3}\left\langle N^{xs}(w): (\lambda\gamma_v\theta):(z_1):(\lambda\gamma^r\theta)(\theta\gamma^{tuv}\theta):(z_2) :(\lambda\gamma^n\theta):(z_3)\right\rangle\non\\
&=&-\f{\alpha'}{4}\oint_{z_1}\f{dw}{w-z_3}\left\langle:\left(\f{(\gamma^{xs})^\alpha_{\;\;\sigma}\lambda^\sigma(\gamma_v\theta)_\alpha(z_1) }{w-z_1}  \right): :(\lambda\gamma^r\theta)(\theta\gamma^{tuv}\theta):(z_2): (\lambda\gamma^n\theta):(z_3)\right\rangle\non\\
&&-\f{\alpha'}{4}\oint_{z_2}\f{dw}{w-z_3}\left\langle:(\lambda\gamma_v\theta)(z_1)::\left(\f{(\gamma^{xs})^\alpha_{\;\;\sigma}\lambda^\sigma(\gamma^r\theta)_\alpha(z_2)}{w-z_2} \right)(\theta\gamma^{tuv}\theta)(z_2) :: (\lambda\gamma^n\theta)(z_3):\right\rangle\non\\
&=&\f{\alpha'}{4(z_1-z_3)}\left\langle:(\lambda \gamma^{xs} \gamma_v\theta):(z_1) :(\lambda\gamma^r\theta)(\theta\gamma^{tuv}\theta):(z_2) (\lambda\gamma^n\theta)(z_3)\right\rangle\non\\
&&+\f{\alpha'}{4(z_2-z_3)}\left\langle:(\lambda\gamma_v\theta):(z_1):(\lambda \gamma^{xs} \gamma^r\theta)(\theta\gamma^{tuv}\theta):(z_2) : (\lambda\gamma^n\theta):(z_3)\right\rangle\non
\ee
The first correlator in \eqref{5.8} can be similarly worked out. Thus, we obtain on using the pure spinor correlators given in appendix \ref{PureSuperspace}, the momentum conservation and the transversality of the polarization tensors
\be
I'&=&\f{i}{5120}e^{mnp}e^{(1)}_{p}e^{(2)}_{n}(p_2)_m
\ee
The contribution of other terms in the expression of $N^{mn}\lambda^\alpha F_{\alpha mn}$ can be similarly worked out. Their total contribution is given by
\be
-\f{i}{8192}e^{mnp}e_{p}^{(1)}e_{n}^{(2)}(p_2)_m\label{2nd_contri}
\ee 
To obtain the full contribution due to the terms considered here, we also need to add to the contributions of \eqref{1st_contri} and \eqref{2nd_contri}, the contribution obtained by interchanging the first massless vertex with the second massless vertex (i.e. $1\leftrightarrow 2$). In our case, this simply doubles the result. 

\vspace*{.06in} As mentioned earlier, the total contribution to the amplitude also receives contributions coming from the higher orders in $\theta$ 
expansion.  For this amplitude, since there is only one kinematic structure possible\footnote{One can easily see this by trying to construct Lorentz 
invariant kinematic structures using the momenta and polarization tensors, taking into account the momentum conservation and the transversality conditions 
\eqref{trans_condi}}, we expect the contribution of $d_{\alpha}\lambda^\beta C^\alpha_{\;\;\beta}$ term in the massive vertex operator to yield the same kinematic structure as in \eqref{1st_contri}
and \eqref{2nd_contri} and hence providing additional contribution to the numerical factors of these equations.

\vspace*{.06in}A priori, the overall numerical factors of pure spinor and 
RNS amplitudes will not agree. The comparison between the numerical factors of two results will fix the relative normalization between them. With this 
normalization, all subsequent amplitudes must agree even upto the numerical factors.

\vspace*{.06in}The corresponding amplitude in RNS formalism is straightforward to compute (see, e.g. \cite{Park} and references therein). 
The kinematic structure of the contribution to the $\langle AAb\rangle$ amplitude, considered partially above, which is computed using the pure spinor 
formalism matches with the corresponding kinematic structure computed using the RNS formalism\footnote{\cite{Park} also got the same kinematic structure 
using pure spinor. This is not surprising since there is a unique kinematic structure for this amplitude. However, the numerical prefactor crucially depends
on the $\theta$ expansion of vertex operator (which was not given in \cite{Park}). \label{f3}}.

\sectiono{Conclusion} \label{discus}
We have given the systematic procedure to obtain the $\theta$ expansion of first massive vertex operator in the pure spinor formalism of superstring theory. 
This result is essential for computing the scattering amplitudes involving the massive states in pure spinor. We have also given the explicit $\theta$ 
expansion of the superfields appearing in this vertex operator upto $O(\theta^3)$. The higher $\theta$ components can be straightforwardly computed using 
the procedure described in this paper. With the help of these and the corresponding result for the superfields describing the massless states, 
any tree level amplitude with three or less number of first massive states and arbitrary number of massless states can now be computed in pure spinor formalism.
Similarly, one loop amplitudes which involve just one massive state can also be computed.
The 4 and higher point tree amplitudes involving more than three massive states and two and higher loop amplitudes involving any number of massive states 
require the integrated form of the vertex operator \cite{next}. All these calculations are tedious and prone to error when performed manually. Hence, it is more 
efficient to use a computer code to perform the $\theta$ expansion and for doing the amplitude calculations \cite{code_paper}.

\vspace*{0.06in} This construction can be readily extended to obtain the first massive vertex operator in heterotic and type II superstring theories in the pure spinor formalism. For heterotic superstring, we simply take the tensor product of the vertex given in this paper with the anti-holomorphic vertex of the bosonic string. Whereas for the type II theories, we take the   tensor product of the holomorphic and anti-holomorphic copies of the vertex given here.

\vspace*{.06in}Along the way, we have also found the relations, hitherto unknown,  which are obeyed by the basic superfields $B_{mnp}, G_{mn}$ and $\Psi_{m\alpha}$ which describe the massive spin-2  supermultiplet.  These relations  are part of superspace description of massive spin-2 multiplet. Furthermore, they are necessary to ensure that we have the correct physical degrees of freedom and are pivotal to perform $\theta$ expansion.

\bigskip

\noindent{\bf Acknowledgments:}  
We are deeply thankful to Ashoke Sen for suggesting to look into the problem, for numerous illuminating discussions throughout the course of this work and for very insightful comments on the draft. We would also like to thank Anirban Basu, Rajesh Gopakumar and Satchitananda Naik for their encouragement and comments on the draft. The work of MV was also supported by the SPM fellowship of CSIR. We also thank the people and Government of India for their continuous support for theoretical physics. 
\appendix

\sectiono{Summary of conventions} \label{saconv}

In this appendix, we give a summary of the notations and
conventions we have used in this paper.

\begin{itemize}

\item Our (anti)symmetrization convention is as follows
\be
\text{Anti-symmetrization}\quad:\qquad  T^{[m_1...m_n]} \equiv \frac{1}{n!} (T^{m_1...m_n} \pm \text{all\; permutations })
\ee 
\be
\text{Symmetrization}\quad:\qquad T^{(m_1...m_n)} \equiv \frac{1}{n!} (T^{m_1...m_n} + \text{all\; permutations})
\ee
\item All antisymmetric products of gamma matrices are defined as
\be
\gamma^{m_1...m_p} \equiv \gamma^{[m_1...}\gamma^{\m_p]} 
\ee
Anti-symmetrized product of $p$ gamma matrices is sometimes referred to as $p$-form. 
\item Our convention for super-covariant derivative is
\be
D_\alpha = \partial_\alpha +  (\gamma^{m})_{\alpha \beta} \theta^{\beta} \partial_{m}\, ; \; \text{where}\quad \partial_\alpha \equiv \frac{\partial}{\partial \theta^\alpha}
\ee
Therefore, the Clifford identity of gamma matrices implies
\be
\lbrace D_\alpha, D_\beta\rbrace= 2 (\gamma^m)_{\alpha \beta} \partial_m \quad\implies\qquad (\gamma_m)^{\alpha\beta}D_\alpha D_\beta = \f{1}{16}\p_m
\ee
In momentum space, this implies for the first massive state
\be
 k^m(\gamma_m)^{\alpha\beta}D_\alpha D_\beta = \f{i}{16}k^mk_m=-\f{i}{16\alpha'}
\label{A.5}
\ee
\item All normal ordering of products of operators are considered to be generalized normal ordering defined as follows-
\be 
:AB:(z) \equiv \frac{1}{2 \pi i} \oint_z  \frac{dw}{w-z} A(w)B(z)  \; , \; \;\text{   For any two operators $A$ and $B$.}
\ee

\end{itemize}

\sectiono{Useful identities involving gamma matrices in d=10}  \label{identi}

In this appendix, we write down the list of gamma matrix identities that were used in our calculations. A useful reference for the more exhaustive list is \cite{Guttenberg}. Most of the manipulations involving the gamma matrices were done with the help of the Mathematica package Gamma \cite{Gamma}.

\vspace*{.06in}We work solely with $16 \times 16$ gamma matrices in $d =10$. These are the off-diagonal elements of the $32\times 32$ gamma matrices
$\Gamma^{m}$ matrices
satisfying 
\be 
\{\Gamma^{m},\Gamma^{n}\}=2\eta^{mn}\mathbb{I}_{32\times 32}\non
\ee
More specifically, 
\be\Gamma^{m}=\begin{pmatrix}
0 & (\gamma^{m})_{\alpha\beta}\\
(\gamma^{m})^{\alpha\beta}& 0\non
\end{pmatrix}
\ee

\begin{itemize}
\item \textbf{Spinor index structure of various gamma matrices}\\
Following is the spinor index structure for various antisymmetric products of gamma matrices
\be
(\gamma^{m_1...m_n})^{\alpha}_{\; \; \beta} \qquad \text{or}\qquad (\gamma^{m_1...m_n})_{ \beta}^{\; \;\alpha}\qquad \text{for $n=0,2,4,6,8,10$}\non
\ee
\be
(\gamma^{m_1...m_n})_{\alpha \beta} \qquad\text{or } \qquad(\gamma^{m_1...m_n})^{\alpha \beta}  \qquad\text{for $n=1,3,5,7,9$}\non
\ee
\item \textbf{Hodge duals}

For $10$ dimensional $16\times16$ gamma matrices, the hodge duality is more than mere duality. It turns out to be an equality. We summarize them below 
\be 
(\gamma^{m_1...m_{2n}})^{\alpha}_{\; \; \beta} &=& \frac{1}{(10-2n)!}(-1)^{(n+1)} \epsilon^{m_1...m_{2n} p_1...p_{10-2n}}(\gamma_{p_1...p_{10-2n}})^{\alpha}_{\; \; \beta}
\ee
\be 
(\gamma^{m_1...m_{2n}})_{\alpha}^{\; \; \beta} &=& -\frac{1}{(10-2n)!}(-1)^{(n+1)} \epsilon^{m_1...m_{2n} p_1...p_{10-2n}}(\gamma_{p_1...p_{10-2n}})_{\alpha}^{\; \; \beta}
\ee
\be 
(\gamma^{m_1...m_{2n+1}})^{\alpha \beta} &=& \frac{1}{(9-2n)!}(-1)^{n} \epsilon^{m_1...m_{2n+1} p_1...p_{9-2n}}(\gamma_{p_1...p_{9-2n}})^{\alpha \beta}
\ee
\be 
(\gamma^{m_1...m_{2n}})_{\alpha \beta} &=& -\frac{1}{(9-2n)!}(-1)^{n} \epsilon^{m_1...m_{2n+1} p_1...p_{9-2n}}(\gamma_{p_1...p_{9-2n}})_{\alpha \beta}
\ee
where, $\epsilon^{m_1\cdots m_{9}}$ is the 10 dimensional epsilon tensor defined as
\be
\epsilon_{0\;1\;\cdots\;9}=1\qquad\implies \qquad \epsilon^{0\;1\;\cdots\;9}=-1
\ee
Due to the above dualities, we only take $\gamma^{m_1}$, $\gamma^{m_1 m_2}$, $\gamma^{m_1 m_2 m_3}$, $\gamma^{m_1 m_2 m_3 m_4 }$ and $\gamma^{m_1 m_2 m_3 m_4 m_5}$ 
along with the identity matrix $\mathbb{I}_{16 \times 16}$ as the linearly independent basis elements for vector spaces of $16 \times 16$ complex matrices. 

\item \textbf{Symmetry property of gamma matrices under exchange of Spinor indices}

\be 
(\gamma^m)_{\alpha \beta} &=& (\gamma^m)_{\beta \alpha} \qquad \text{: Symmetric  }
\ee
\be 
(\gamma^{m_1 m_2})^{\alpha}_{\; \; \beta} &=& -(\gamma^{m_1 m_2})_{\beta}^{\; \; \alpha} \qquad\text{: Anti-Symmetric}
\ee
\be 
(\gamma^{m_1 m_2 m_3})_{\alpha \beta} &=& -(\gamma^{m_1 m_2 m_3})_{\beta \alpha} \qquad \text{: Anti-Symmetric}
\ee
\be 
(\gamma^{m_1 m_2 m_3 m_4})^{\alpha}_{\; \; \beta} &=& (\gamma^{m_1 m_2 m_3 m_4})_{\beta}^{\; \; \alpha} \qquad\text{: Symmetric}
\ee
\be 
(\gamma^{m_1 m_2 m_3 m_4 m_5})_{\alpha \beta} &=& (\gamma^{m_1 m_2 m_3 m_4 m_5})_{\beta \alpha} \qquad \text{: Symmetric}
\ee
For 1, 3 and 5 forms, the same (anti) symmetry properties hold when the spinor indices are upstairs.  

\item \textbf{Various Gamma Traces}
\be
(\gamma^{m_1...m_n})^{\alpha}_{\;\,\alpha} &=& 0 \quad
\text{for} \quad n=2,4,6,8
\ee
\be
(\gamma^{m_1...m_{10}})^{\alpha}_{\;\,\alpha} &=& -16\; \epsilon^{m_1...m_{10}} 
\ee
\be 
(\gamma^m)_{\alpha \beta} (\gamma_n)^{ \beta\alpha}&=& 16\; \delta^m_n
\ee
\be
(\gamma^{m_1...m_n})_{\alpha \beta}(\gamma_{p_n...p_1})^{\beta \alpha}& =& 16n! \;\delta^{m_1...m_n}_{p_1...p_n} ,\;\text{for $n$ $\in$ odd}
\ee
\be
(\gamma^{m_1...m_n})^{\alpha}_{\; \; \beta}(\gamma_{p_n...p_1})^{\beta}_{\; \; \alpha} &=& 16n!\; \delta^{m_1...m_n}_{p_1...p_n} ,\;\text{for $n$ $\in$ even}
\ee

\item \textbf{Bi-Spinor decomposition}

Any Bi-spinor $T_{\alpha\beta}$ can be decomposed as  
\be 
T_{\alpha \beta} = t_\m (\gamma^m)_{\alpha \beta} + t_{m n p} (\gamma^{m n p})_{\alpha \beta}+ t_{m n p q r} (\gamma^{m n p q r})_{\alpha \beta} 
\label{bispinor_expansion}
\ee
where, for $r=1,3,5$
\be
  t_{m_1...m_r}= \frac{1}{16r!}(\gamma_{m_1...m_r})^{\alpha \beta}\; T_{\alpha \beta}
  \ee
Similarly, a tensor-spinor $T^\alpha_{\;\;\beta}$ can be decomposed as
\be 
T^{\alpha}_{\;\,\beta} = t \, \delta^{\alpha}_{\;\,\beta} + t_{m n} (\gamma^{m n })^{\alpha}_{\;\,\beta}+ t_{m n p q} (\gamma^{m n p q})^{\alpha}_{\;\,\beta} 
\ee
where, for $r = 2,4$
\be
 t_{m_1...m_r}= \frac{1}{16r!}(\gamma_{m_1...m_r})^{\alpha}_{\;\,\beta}\; T^{\alpha}_{\;\,\beta}
\ee

\item \textbf{Tensor index contracted identities involving gamma matrices}
\be 
(\gamma^{m n})^{\alpha}_{\; \; \beta}(\gamma_{m n})^{\rho}_{\; \; \lambda} = 4 (\gamma^m)_{\beta \lambda} (\gamma_m)^{\alpha \rho} - 2 \delta^{\alpha}_{\beta} \delta^{\rho}_{\lambda}-8 \delta^{\alpha}_{\lambda}\delta^{\rho}_{\beta}
\ee
\be 
(\gamma^{m n})^{\alpha}_{\; \; \beta} (\gamma_{m n p})^{\rho \lambda} = 2 (\gamma^m)^{\alpha \rho} (\gamma_{p m})^{\lambda}_{\; \; \beta} + 6 (\gamma_p)^{\alpha \rho} \delta^{\lambda}_{\beta} - (\rho \leftrightarrow \lambda )
\ee
\be 
(\gamma_{m n})^{\alpha}_{\; \; \beta} (\gamma^{m n p})_{\rho \lambda} = -2 (\gamma_m)_{\beta \lambda} (\gamma^{p m})^{\alpha}_{\; \;\rho} + 6 (\gamma^p)_{\beta \lambda} \delta^{\alpha}_{\rho} - (\rho \leftrightarrow \lambda )
\ee
\be 
(\gamma_{m n p})^{\alpha \beta}(\gamma^{m n p})^{\rho\lambda} = 12 [(\gamma_m)^{\alpha \lambda}(\gamma^m)^{\beta \rho}-(\gamma_m)^{\alpha \rho}(\gamma^m)^{\beta \lambda}]
\ee
\be 
(\gamma_{m n p})^{\alpha \beta}(\gamma^{m n p})_{\rho\lambda} = 48 (\delta^{\alpha}_{\rho}\delta^{\beta}_{\lambda}-\delta^{\alpha}_{\lambda}\delta^{\beta}_{\rho})
\ee
\end{itemize}

\sectiono{Pure spinor superspace identities}  \label{PureSuperspace}
The pure spinor superspace identities which are used in this paper are listed below \cite{Berkovits_Mafra}
\be
\langle(\lambda\gamma^m\theta)(\lambda\gamma^n\theta)(\lambda\gamma^p\theta)(\theta\gamma_{stu}\theta)\rangle=\f{1}{120}\delta^{mnp}_{stu}
\ee
\be
\langle(\lambda\gamma^{pqr}\theta)(\lambda\gamma_m\theta)(\lambda\gamma_n\theta)(\theta\gamma_{stu}\theta)\rangle=\f{1}{70}\delta^{[p}_{[m}\eta_{n][s}\delta^q_{t}\delta^{r]}_{u]}\label{C.2}
\ee
\be
\langle(\lambda\gamma^{mnpqr}\theta)(\lambda\gamma_s\theta)(\lambda\gamma_{t}\theta)(\theta\gamma_{uvw}\theta)\rangle=-\f{1}{42}\delta^{mnpqr}_{stuvw}-\f{1}{5040}\epsilon^{mnpqr}_{\;\;\;\;\;\;\;\;\;\;\;\; stuvw}\label{C.3}
\ee

\be
\langle(\lambda\gamma_{q}\theta)(\lambda\gamma^{mnp}\theta)(\lambda\gamma^{rst}\theta)(\theta\gamma_{uvw}\theta)\rangle
&=&-\f{1}{280}\Bigl[\eta_{q[u}\eta^{z[r}\delta^{s}_{v}\eta^{t][m}\delta^n_{w]}\delta^{p]}_{z}-\eta_{q[u}\eta^{z[m}\delta^{n}_{v}\eta^{p][r}\delta^s_{w]}\delta^{t]}_{z}\Bigl]\non\\
&&+\f{1}{140}\Bigl[\delta^{[m}_{q}\delta^{n}_{[u}\eta^{p][r}\delta^s_{v}\delta^{t]}_{w]}-\delta^{[r}_{q}\delta^{s}_{[u}\eta^{t][m}\delta^n_{v}\delta^{p]}_{w]}\Bigl]\non\\
&&-\f{1}{8400}\epsilon^{qmnprstuvw}
\ee
\be
&&\hspace*{-.95in}\langle(\lambda\gamma^{mnpqr}\theta)(\lambda\gamma_{stu}\theta)(\lambda\gamma^{v}\theta)(\theta\gamma_{wxy}\theta)\rangle\non\\
&=&\f{1}{120}\epsilon^{mnpqr}_{\;\;\;\;\;\;\;\;\;\;\;\;ghijk}\left(\f{1}{35}\eta^{v[g}\delta^{h}_{[s}\delta^{i}_{t}\eta_{u][w}\delta^j_{x}\delta^{k]}_{y]}-\f{2}{35}\delta^{[g}_{[s}\delta^{h}_{t}\delta^{i}_{u]}\delta^{j}_{[w}\delta^{k]}_{x}\delta^{v}_{y]}\right)\non\\
&&+\f{1}{35}\eta^{v[m}\delta^{n}_{[s}\delta^{p}_{t}\eta_{u][w}\delta^q_{x}\delta^{r]}_{y]}-\f{2}{35}\delta^{[m}_{[s}\delta^{n}_{t}\delta^{p}_{u]}\delta^{q}_{[w}\delta^{r]}_{x}\delta^{v}_{y]}
\ee
\sectiono{Massless vertex operator}
\label{sec:massless}
The massless spectrum contains the physical fields gluon and gluino denoted by $a_m(x)=e_m e^{ik\cdot x}$ and $\xi^\alpha(x)=\chi^\alpha e^{ik\cdot x}$ respectively. Each of them have 8 on-shell degrees of freedom.  
The pure spinor unintegrated $(V)$ and integrated $(U)$ vertex operators  describing these fields are given by
\be
V&=&\lambda^\alpha A_\alpha\non\\[.15in]
U&=&\p\theta^\alpha A_\alpha+\Pi^mA_m+d_\alpha W^\alpha+\f{1}{2}N^{mn}\mathcal{F}_{mn}\non
\ee
The $\theta$ expansion of various fields appearing in the vertex operators are \cite{Harnad, Ooguri, Policastro}
\be
A_\alpha(X,\theta)&=&\f{1}{2}a_m(\gamma^m\theta)_\alpha-\f{1}{3}(\xi\gamma_m\theta)(\gamma^m\theta)_\alpha
-\f{1}{32}F_{mn}(\gamma_p\theta)_\alpha(\theta\gamma^{mnp}\theta)\non\\
&&+\f{1}{60}(\gamma^m\theta)_\alpha (\theta\gamma^{mnp}\theta)(\p_n\xi\gamma_p\theta) +\f{1}{1152}(\gamma^m\theta)_\alpha(\theta\gamma^{mrs}\theta)(\theta\gamma^{spq}\theta)\p_rF_{pq}+\cdots\non\\[.2in]
A_m(X,\theta)&=&a_m-(\xi\gamma_m\theta)-\f{1}{8}(\theta\gamma_m\gamma^{pq}\theta)F_{pq}
+\f{1}{12}(\theta\gamma_m\gamma^{pq}\theta)(\p_p\xi\gamma_q\theta)\non\\
&&+\f{1}{192}(\theta\gamma^{mrs}\theta)_\alpha (\theta\gamma^{spq}\theta)(\p_rF_{pq}) +\cdots\non\\[.2in]
\mathcal{F}_{mn}(X,\theta)&=&F_{mn}-2(\p_{[m}\xi\gamma_{n]}\theta)+\f{1}{4}(\theta\gamma_{[m}\gamma^{pq}\theta)\p_{n]}F_{pq}
-\f{1}{6}(\theta\gamma_{[m}\gamma^{pq}\theta)\p_{n]}\p_p(\xi\gamma_q\theta)\non\\
&&-\f{1}{96}(\theta\gamma_{[m}\gamma^{rs}\theta) (\theta\gamma^{spq}\theta)\p_{n]}\p_rF_{pq} +\cdots\non\\[.2in]
W^\alpha(X,\theta)&=&\xi^\alpha-\f{1}{4}(\gamma^{mn}\theta)^\alpha F_{mn}+\f{1}{4}(\gamma^{mn}\theta)^\alpha (\p_m\xi\gamma_n\theta)+\f{1}{48}(\gamma^{mn}\theta)^\alpha(\theta\gamma_n\gamma^{pq}\theta)\p_mF_{pq}
\non\\
&&-\f{1}{96}(\gamma^{mn}\theta)^\alpha (\theta\gamma^{npq}\theta)\p_m\p_p(\xi\gamma_q\theta) -\f{1}{1536}(\gamma^{mn}\theta)^\alpha(\theta\gamma^{nrs}\theta)(\theta\gamma^{spq}\theta)\p_r\p_mF_{pq}+\cdots\non
\ee
where, $F_{mn}=2\p_{[m}a_{n]}$ describes the gluon field strength.

\sectiono{Consistency of the differential relations with $QV=0$ }
\label{sec:check}
In this appendix we give an outline of the proof that the relations given in section \ref{ingredients} are consistent with the equations of motion \eqref{1st} to \eqref{7th} and among themselves. We indicate the steps for the consistency with equations \eqref{1st} and \eqref{2nd}. The rest of the equations can be verified in a similar fashion. 

\vspace*{.06in}We first show that our expression for $D_\alpha B_{mnp}$ is consistent with the two expressions of $H_{s\beta}$ given in  equations \eqref{2.14} and \eqref{3.21}. For this, we note that by putting the expression of $D_\alpha B_{mnp}$ from \eqref{D_Bmnp_general} into the expression of $H_{s\alpha}$ given in \eqref{2.14}, we obtain
\be
H_{s\alpha}&=&\f{3}{7}(\gamma^{mn})_\alpha^{\;\;\beta}D_\beta B_{mns}\non\\
&=&\f{3}{7}(\gamma^{mn})_\alpha^{\;\;\beta}\Bigl[12(\gamma_{[mn}\Psi_{p]})_{\beta}+24\alpha'k^t k_{[m}(\gamma_{|t|n}\Psi_{p]})_{\beta} \Bigl]\non\\
&=&-96\Psi_{s\alpha}+24\Psi_{s\alpha}\non\\
&=&-72\Psi_{s\alpha}\non
\ee
In the rest frame, this matches with the corresponding result given in \eqref{2.20}. If we only consider the momentum independent term in the covariant expression of $D_\alpha B_{mnp}$ as done in \cite{Park}, then the two expressions of $H_{s\alpha}$ do not match with each other (see also footnotes \ref{f1} and \ref{f2}). 

\vspace*{.06in} Next, on substituting \eqref{3.21} and \eqref{D_Bmnp_general}  in  \eqref{1st} and using \eqref{KG11} along with $k^m\Psi_{m\alpha}=0$ of \eqref{cons_theta=0}, we obtain 
\be
LHS&=&(\gamma_{mnpqr})^{\alpha\beta} \left[D_\alpha B_{\beta\sigma}-\gamma^s_{\alpha\sigma}H_{s\beta}\right]\non\\
&=&-12(\gamma^{stu}\gamma_{mnpqr}\gamma_{st})^{\;\;\alpha}_{\sigma}\Psi_{u\alpha}-24\alpha'k^vk_s(\gamma^{stu}\gamma_{mnpqr}\gamma_{vt})^{\;\;\alpha}_{\sigma}\Psi_{u\alpha} +72 (\gamma^s\gamma_{mnpqr})^{\;\;\beta}_{\sigma}\Psi_{s\beta}\non\\
&=&0\non
\ee
This shows that the equation \eqref{1st} is identically satisfied.

\vspace*{.06in}Next, we consider equation \eqref{2nd}. Using the expression of $C_{mnpq}$ from \eqref{2.14}, expression of $H_{s\beta}$ from equation \eqref{3.21}, equation \eqref{D_Psi1} and noting that the trace of product of 5 form with 1 and 3 form is zero, we obtain for the left hand side of \eqref{2nd}
\be
LHS&=& (\gamma_{mnpqr})^{\alpha\beta} \left[D_\alpha H_{s\beta}-(\gamma_s)_{\alpha\sigma}C^\sigma_{\;\;\beta}\right]\non\\
&=&\f{i}{2}(\gamma_{mnpqr}\gamma_{stuvw})^\beta_{\;\;\beta}k^tB^{uvw}-\f{i}{2}(\gamma_{mnpqr}\gamma_s\gamma^{tuvw})^\beta_{\;\;\beta}k_tB_{uvw}\non\\
&=&0\non
\ee
Hence, \eqref{2nd} is also identically satisfied.

\vspace*{.06in}To verify the rest of the equations resulting from $QV=0$, one can follow similar steps as in the above two cases. All one needs to use are the various gamma matrix identities and equations \eqref{2.14} - \eqref{KG11}, \eqref{3.21} and \eqref{D_Gmn1} - \eqref{cons_theta=0}. Using these, we can show that all the remaining equations, viz. \eqref{Berkovits_eqns} to \eqref{7th} are satisfied identically. This establishes the consistency of our proposed relations in section \ref{ingredients} with $QV=0$.

\vspace*{.06in}Finally, we also need to verify that all the relations given in section \ref{ingredients} are consistent with each other. One way to verify this is to take the supercovariant derivative of both sides of equations \eqref{D_Gmn1}-\eqref{D_Psi1}. Using the identity \eqref{A.5}, the left hand side of these equations will become proportional to a single superfield, whereas the right hand sides will now involve the supercovariant derivatives of the superfields. The RHS can be shown to be identical to LHS using equations \eqref{D_Gmn1} - \eqref{cons_theta=0} and various gamma matrix identities. The consistency of \eqref{cons_theta=0} with the equations \eqref{D_Gmn1} - \eqref{D_Psi1} is also easy to verify. The consistency of our proposed relations with $QV=0$ and among themselves is therefore established.

\end{document}